# A novel explicit design method for complex thin-walled structures based on embedded solid moving morphable components

Wendong Huo[1], Chang Liu[1,2*], Yunpu Liu[1], Zongliang Du[1,2], Weisheng Zhang[1,2], Xu Guo[1,2*]

[1]*State Key Laboratory of Structural Analysis, Optimization and CAE Software for Industrial Equipment,*
*Department of Engineering Mechanics,*
*International Research Center for Computational Mechanics,*
*Dalian University of Technology, Dalian, 116023, P.R. China*

[2]*Ningbo Institute of Dalian University of Technology, Ningbo, 315016, P.R. China*

## Abstract

In this article, a novel explicit approach for designing complex thin-walled structures based on the Moving Morphable Component (MMC) method is proposed, which provides a unified framework to systematically address various design issues, including topology optimization, reinforced-rib layout optimization, and sandwich structure design problems. The complexity of thin-walled structures mainly comes from flexible geometries and the variation of thickness. On the one hand, the geometric complexity of thin-walled structures leads to the difficulty in automatically describing material distribution (e.g., reinforced ribs). On the other hand, thin-walled structures with different thicknesses require various hypotheses (e.g., Kirchhoff-Love shell theory and Reissner-Mindlin shell theory) to ensure the precision of structural responses. Whereas for cases that do not fit the shell hypothesis, the precision loss of response solutions is nonnegligible in the optimization process since the accumulation of errors will cause

* Corresponding authors. E-mail: c.liu@dlut.edu.cn (Chang Liu), guoxu@dlut.edu.cn (Xu Guo)

entirely different designs. Hence, the current article proposes a novel embedded solid component to tackle these challenges. The geometric constraints that make the components fit to the curved thin-walled structure are whereby satisfied. Compared with traditional strategies, the proposed method is free from the limit of shell assumptions of structural analysis and can achieve optimized designs with clear load transmission paths at the cost of few design variables and degrees of freedom for finite element analysis (FEA). Finally, we apply the proposed method to several representative examples to demonstrate its effectiveness, efficiency, versatility, and potential to handle complex industrial structures.

## 1. Introduction

Thin-walled structure has a pivotal role in engineering applications, such a structure form is widely employed as the main load-carrying structures in heavy equipment, such as vessels, airplanes, and spacecraft. Although thin and light, thin-walled structures span over relatively large areas and hold externally applied load in an efficient way [1]. Whereas, thin-walled structures commonly suffer from extreme loads and complex service environments in practical applications. Hence, topology optimization on thin-walled structures is widely investigated to enhance mechanical properties and further resist deformation, buckling, and vibration.

Topology optimization aims at finding the optimal material distribution in the prescribed design domain under various constraints. With the aid of such a powerful tool, engineers can find inspiring structural configurations and create competitive products systematically. The invention of this revolutionary design tool can trace back to the pioneering work of Bendsøe and Kikuchi [2]. Since then, many topology optimization methods, including the Solid Isotropic Material with Penalization (SIMP) method [3, 4], evolutionary structural optimization (ESO) method [5], and level set method [6, 7], have been developed and successfully applied in numerous scientific and engineering applications.

Different from traditional topology optimization problems, the main difficulties and challenges in thin-walled structure design are as follows. First, the shape of the concerned thin-walled structure is usually flexible due to the complexity of geometry, which may involve complicated surface operations to form a structure with clear boundaries. Secondly, the finite element meshes must be relatively refined to approximate the complex geometry [8], which will increase the number of design variables tremendously under the framework of implicit optimization methods. Last but not least, the filter approaches, which are very effective in restraining the numerical instabilities (e.g., islanding effect, grey elements, and the checkerboard pattern) [9, 10] of the pixel-based topology optimization method, are originally developed in flat space (2D or 3D) and may encounter some challenges when applying on thin-walled structures [11] due to the variation of surface curvature.

On account of the abovementioned challenges, the field of structural design on thin-walled structures has attracted attention in recent years. Liu et al. [12] proposed the Heaviside-function-based directional growth topology parameterization (H-DGTP) of the casting constraints and successfully applied it for simultaneously optimizing the layout and size of the stiffeners on thin-walled structures. Based on the smeared stiffener method (SSM), Hao et al. [13] developed a hybrid minimum-weight optimization formulation for hierarchical stiffened thin-walled structures against buckling loads. Later, Wang et al. [14] extended the SSM method to the NSSM method. A new efficient and accurate numerical solution technique for large-scale stiffened shells is introduced to overcome the drawbacks of the traditional SSM method. Inspired by natural structures, such as plant leaves, nut shells, and insect wings, Wang et al. [15] proposed a bio-inspired conceptual design method for composite thin-walled structures with stiffeners. Li et al. [16] developed an evolutionary design algorithm to solve the optimal stiffener layout problem for thin-walled structures to maximize stiffness with volume constrained. Träff et al. [17] developed a high-performance computing framework for solving ultra-large-scale thin-walled structure topology optimization problems, where the multigrid approach with good parallel scaling properties is utilized to solve the linear system. Deng et al. [11] introduced the filter formulation constructed on surfaces to alleviate numerical instabilities in topology optimization on thin-walled structures. Some other recent excellent progress can be found in references [18-27].

On the other hand, the mapping-based approaches also received a lot of attention recently, owing to their capability of handling complex geometries. Hassani et al. [28] realized simultaneously optimizing the shape and topology of thin-walled structures, where the Non-Uniform Rational B-Spline (NURBS) description is employed to model and control geometry shape. Choi et al. [29] proposed a constrained isogeometric design optimization method for thin-walled lattice structures, where the free-form deformation and the global curve interpolation are utilized to define the control net of lattice structure on curved middle surfaces. Constructed upon the embedded isogeometric Kirchhoff-Love shell theory, Hirschler et al. [30] developed a novel approach that deals with rib layout optimization and shape optimization on non-conforming multi-patch thin-walled structures, which is capable of tackling the geometric constraint of connecting interfaces between the panel and the stiffeners. Feng et al. [31] proposed an effective B-spline parameterization method for stiffener layout optimization on thin-walled shell structures, which produces checkerboard-free design results with a clear layout and avoids overhang stiffeners.

Different from the geometric mappings constructed by smooth analytical functions in the spline-type surface descriptions, the mesh parameterization technique establishes a numerical mapping relation from a mesh surface to a planar parametric domain, where the surface can be obtained from a point cloud, a CAD model or a mesh model. Hence, mesh parameterization has a lower construction cost and receives growing attention in the field of designing thin-walled structures. Based on the so-called ARAP (As-Rigid-As Possible) parameterization technique and the cutting operations, Zhang et al. [32] realized the stiffener layout optimization on thin-walled structures under the framework of topology optimization, where the stiffeners are parameterized through B-splines, and the given 3D surface is mapped to the related parametric domain via mesh parameterization. Ye et al. [33] first introduced the conformal parameterization technique to topology optimization and further developed a novel level-set-based method for systematically optimizing the structural topology of thin-walled shells. Later on, the concept of the surface level set method is extended to the field of ferromagnetic soft active thin-walled structure design [34] and thermal control structure design [35].

Although the above works have led to significant breakthroughs, the major drawback of these methods is that they are mainly constructed based on implicit topology optimization approaches, and may cause problems such as the islanding effect, grey elements, checkboard phenomenon, and enormous design variables. In the last decade, explicit topology optimization methods have received considerable attention and rapid development [36-38]. Among them, the Moving Morphable Component (MMC) method uses a set of components with explicit geometric descriptions to define the structural topology [36]. Benefiting from explicit geometric descriptions, the MMC method require no filtering operators, have few design variables, and can obtain clear load paths. Currently, the MMC method has been extended to various scientific and engineering application scenarios, including minimum length scale control [39], additive manufacturing [40], structural dynamic responses [41], and multi-physics problems [42], etc. In recent years, there has been a growing trend of research applying explicit topology optimization methods to the field of thin-walled structure design [43-50].

However, several problems with the explicit topology optimization of thin-walled structures still have not been well addressed. One is that under shell-based solution frameworks, the displacements along the thickness direction are normally considered as varying linearly, which inevitably reduces the accuracy of response solutions. Whereas the loss of precision is nonnegligible in the optimization process considering that the accumulation of errors will cause entirely different designs. On the other hand, the shell model has difficulty in describing different design problems unitedly, e.g., topology optimization, stiffener layout optimization, and sandwich structure optimization. Considering these above issues, the present article intends to propose a novel embedded component description based on the Computational Conformal

Mapping (CCM) technique, in which the embedded component perfectly follows the shape variation of thin-walled structures (shown in Fig. 1).

The remainder of this article is organized as follows. Section 2 mainly introduces the theoretical foundations of the proposed method, including the MMC method, the CCM technique, the topology description function (TDF) construction on complex surfaces, and the description of embedded components. In section 3, problem formulation and sensitivity analysis are derived in detail. Section 4 is devoted to providing specific solution implementations, including the offset-based solid mesh generation scheme, the solution of structural responses, the discrete sensitivities, and the DOF removal technique. Section 5 describes the outline of the primary optimization procedures. In section 6, representative numerical examples are considered to demonstrate the effectiveness and the efficiency of the current algorithm, and some dominating factors/parameters in the thin-walled structure design are investigated. Some conclusions of the proposed algorithm and potential extensions bring the article to a close in section 7.

## 2. Theoretical foundations

### *2.1 Moving Morphable Component (MMC) method*

In conventional implicit topology optimization approaches, design variables are generally defined by pixel-based elements or nodes, which may encounter difficulties when interpreting the final optimization results. To tackle numerical instabilities (including the islanding effect, grey elements, checkboard phenomenon, etc.) from the source, the MMC method is proposed to solve topology optimization problems geometrically [36]. Under the MMC framework, the structures are constructed based on a group of explicitly described components, which makes the design variables contain more geometric, mechanical, and engineering meanings.

In the MMC method, the topology description function (TDF) is employed to characterize the domain occupied by the components. In the present work, the TDF of whole structure is defined as follows:

$$\begin{cases} \phi^s(\boldsymbol{x}) > 0, & \text{if } \boldsymbol{x} \in \Omega_s, \\ \phi^s(\boldsymbol{x}) = 0, & \text{if } \boldsymbol{x} \in \partial\Omega_s, \\ \phi^s(\boldsymbol{x}) < 0, & \text{if } \boldsymbol{x} \in \Omega \backslash (\Omega_s \cup \partial\Omega_s), \end{cases} \tag{2.1}$$

where $\Omega$ denotes the design domain, $\Omega_s$ represents the region occupied by the designed structure. Under the MMC framework, the designed structure is composed of a group of components. Taking the $i$-th component as an example, the geometry of this component is shown in Fig. 2, and the corresponding TDF is defined as:

$$\phi_i = 1 - \left( \left( \frac{x_i'}{L_i} \right)^6 + \left( \frac{y_i'}{f_i(x_i')} \right)^6 \right)^{1/6}, \tag{2.2}$$

with

$$\begin{pmatrix} x_i' \\ y_i' \end{pmatrix} = \begin{bmatrix} \cos\theta_i & \sin\theta_i \\ -\sin\theta_i & \cos\theta_i \end{bmatrix} \begin{pmatrix} x - x_0^i \\ y - y_0^i \end{pmatrix}, \tag{2.3}$$

$$f_i(x_i') = \frac{t_i^1 + t_i^2 - 2t_i^3}{2L_i^2} (x_i')^2 + \frac{t_i^2 - t_i^1}{2L_i} x_i' + t_i^3, \tag{2.4}$$

where $x_0^i$ and $y_0^i$ are the coordinates of the $i$-th component's central point, and $\theta_i$ denotes its rotational angle with respect to the global Cartesian coordinate system. In Eq. (2.2)-Eq. (2.4), $t_i^1$, $t_i^2$, $t_i^3$, and $L_i$ denote three half-thickness parameters and the half-length parameter of the component, respectively. At the structure level, the TDF $\phi^s$ consists of the TDF of each component via the K-S function as:

$$\phi^s = \max(\phi_1, \dots, \phi_i, \dots, \phi_n), \tag{2.5}$$

where $n$ denotes the total number of components in the design domain. Furthermore, the vector of the design variables regarding the $i$-th component is defined as $\boldsymbol{D}_i = (x_0^i, y_0^i, \theta_i, L_i, t_i^1, t_i^2, t_i^3)$, and the vector of the design variables of the whole structure is assembled as $\boldsymbol{D} = (\boldsymbol{D}_1, \boldsymbol{D}_2, \dots, \boldsymbol{D}_n)$.

### *2.2 Computational Conformal Mapping (CCM)*

Due to the complexity and flexibility of geometry models, the conventional MMC approach cannot be directly employed to solve topology optimization problems on thin-walled structures (the obstacle is explained visually in Fig 1(a)). To address this issue, the present study establishes a mapping between the middle surface of the concerned thin-walled structure and a standard parametric domain.

Mapping a surface with an arbitrary shape to a suitable domain, also termed surface parameterization [53], has long been a central issue in computer industries. Owing to the dimension-reduction property, many tough issues such as texture mapping, surface registration, morphing, and remeshing are simplified [54]. Benefitting from the rapid development of computer technology, powerful tools have been invented and introduced to fulfill such a challenging task [55-58]. Among them, two major classes are authalic (area-preserving) mappings and conformal (angle-preserving) mappings. Since the geometries of objects are commonly described by the mesh models, conformal mapping receives much more attention owing to its property of preserving mesh quality. In the present work, the computational method for solving conformal mappings developed in [59-61] is adopted to achieve an MMC-friendly mapping. The core procedures are summarized as follows.

Let $\boldsymbol{S}$ denote a simply-connected open surface in $\mathbb{R}^3$ ($\boldsymbol{S}$ is the middle surface of the concerned thin-walled structure), the goal of the adopted CCM technique is to compute a conformal mapping $f: \boldsymbol{S} \to \boldsymbol{\mathcal{M}}$, where $\boldsymbol{\mathcal{M}}$ is a rectangular parametric domain defined in $\mathbb{R}^2$. Instead of directly computing the mapping, the current algorithm composes two quasi-conformal mappings with proper Beltrami coefficients, which significantly simplifies and speeds up the computation process. Disk harmonic mapping is utilized as the initialized mapping (i.e., the first quasi-conformal mapping), which is easily obtained by solving a system of partial differential equations via the finite element method [62]. More specifically, let $\boldsymbol{\mathcal{D}}$ denote a unit disk in the complex plane $\mathbb{C}$ (see Fig. 3(a)). The harmonic mapping $h: \boldsymbol{S} \to \boldsymbol{\mathcal{D}} \subset \mathbb{C}$ satisfies the following equations:

$$\begin{cases} \Delta_{\boldsymbol{S}} h = 0, \qquad \text{in } \boldsymbol{S}, & (2.6a) \\ h(\partial \boldsymbol{S}) = \partial \boldsymbol{\mathcal{D}}, & (2.6b) \end{cases}$$

where $\Delta_{\boldsymbol{S}}$ represents the Laplace operator defined on surface $\boldsymbol{S}$, $\partial \boldsymbol{S}$ and $\partial \boldsymbol{\mathcal{D}}$ denote the boundaries of $\boldsymbol{S}$ and $\boldsymbol{\mathcal{D}}$, respectively. It's noteworthy that the resultant mapping $h$ has reduced the dimension of the original surface and realized the purpose of surface parameterization, whereas angular distortions are possibly involved in the process. For the sake of revising the conformality distortion (to enhance mesh quality) and achieving the bijectivity (to avoid the overlaps of meshes), the second quasi-conformal mapping is constructed by the so-called linear Beltrami solver (LBS) scheme developed in [59, 60], and the core idea is reconstructing a mapping based on a given Beltrami coefficient.

Let $\mu_{h^{-1}} = \rho + j\tau$ (which is calculated explicitly from the solution of Eq. (2.6)) denote the Beltrami coefficient of the inverse mapping $h^{-1}$, where $\rho$ and $\tau$ are the real and imaginary part, respectively, with $j^2 = -1$ representing the imaginary unit. Suppose $g(z = x + jy) = u(x, y) + jv(x, y): \boldsymbol{\mathcal{D}} \to \boldsymbol{\mathcal{M}}$ is the second desired quasi-conformal mapping (see Fig. 3(b)), whose Beltrami coefficient is given by $\mu_g = \mu_{h^{-1}}$. Here $(x, y)$ and $(u, v)$ are the complex coordinates defined on the unit disk $\boldsymbol{\mathcal{D}}$ and the parameter domain $\boldsymbol{\mathcal{M}}$, respectively. According to the LBS scheme, the components of the mapping $g$ satisfy:

$$\begin{cases} v_y = \alpha_1 u_x + \alpha_2 u_y, & (2.7a) \\ -v_x = \alpha_2 u_x + \alpha_3 u_y, & (2.7b) \end{cases}$$

where $\alpha_1 = \frac{(\rho-1)^2+\tau^2}{1-\rho^2-\tau^2}$, $\alpha_2 = -\frac{2\tau}{1-\rho^2-\tau^2}$, and $\alpha_3 = \frac{(\rho+1)^2+\tau^2}{1-\rho^2-\tau^2}$. Similarly, we have

$$\begin{cases} -u_y = \alpha_1 v_x + \alpha_2 v_y, & (2.8a) \\ u_x = \alpha_2 v_x + \alpha_3 v_y. & (2.8b) \end{cases}$$

Furthermore, since $\left(\frac{\partial(\cdot)}{\partial x}, \frac{\partial(\cdot)}{\partial y}\right) \cdot (v_y, -v_x) = 0$, and $\left(\frac{\partial(\cdot)}{\partial x}, \frac{\partial(\cdot)}{\partial y}\right) \cdot (-u_y, u_x) = 0$, we have

$$\begin{cases} \nabla \cdot \left(\mathbf{A}(\nabla u)\right) = 0, \quad \text{in } \boldsymbol{\mathcal{D}}, & (2.9a) \\ \nabla \cdot \left(\mathbf{A}(\nabla v)\right) = 0, \quad \text{in } \boldsymbol{\mathcal{D}}, & (2.9b) \\ g(\partial \boldsymbol{\mathcal{D}}) = \partial \boldsymbol{\mathcal{M}}, & (2.9c) \end{cases}$$

where $\nabla(\cdot) = \left(\frac{\partial(\cdot)}{\partial x}, \frac{\partial(\cdot)}{\partial y}\right)$, $\mathbf{A} = \begin{pmatrix} \alpha_1 & \alpha_2 \\ \alpha_2 & \alpha_3 \end{pmatrix}$, $\partial \boldsymbol{\mathcal{M}}$ denotes the boundary of the rectangle $\boldsymbol{\mathcal{M}}$. Up to now, the task of constructing the mapping $g$ is simplified to solving the above generalized Laplace equations, which are also solved through the finite element method. Once mappings $h$ and $g$ are obtained, the ultimate mapping $f: \boldsymbol{\mathcal{S}} \to \boldsymbol{\mathcal{M}}$ is established as $f = g \circ h$. The conformality and bijectivity (means no folding or overlaps of meshes exist in the parameterization results) of the acquired mapping are ensured, which are essential for alleviating nonlinearity and avoiding singularity in the problem formulation of structural optimization. For theoretical fundamentals and technical details of the CCM technique, please refer to [59-61] and references therein.

### *2.3 TDF construction on complex surfaces*

In this subsection, the TDF construction on a simply-connected open surface with zero genus is firstly explained. On account of the geometric complexity of a general surface, the current method further integrates the surface-cutting operation and the multi-patch stitching scheme to extend the application scope and enhance the robustness of algorithms.

#### *2.3.1 TDF construction on a simply-connected open surface with zero genus*

Let $\boldsymbol{S}$ denote a simply-connected open surface with zero genus. The parametric domain $\boldsymbol{\mathcal{M}}$ and conformal mapping $f$ are constructed based on the abovementioned CCM technique (as depicted in Fig. 4(a)). The TDF $\phi^{\mathrm{s}}_{\boldsymbol{\mathcal{M}}}(\boldsymbol{p};\boldsymbol{D})$ in parametric domain $\boldsymbol{\mathcal{M}}$ is firstly constructed via the component description of the typical 2D MMC (as shown in Fig. 4(b)), where $\boldsymbol{p}$ and $\boldsymbol{D}$ represent the point in the parametric domain and the vector of design variables, respectively. Since the mapping $f$ is bijective, that is, for each point $\boldsymbol{x} \in \boldsymbol{S}$, there exists a point $f(\boldsymbol{x}) \in \boldsymbol{\mathcal{M}}$ that corresponds to $\boldsymbol{x}$ and vice verse. Based on such a property, the TDF of $\boldsymbol{S}$ is naturally defined as

$$\phi^{\mathrm{s}}_{\boldsymbol{S}}(\boldsymbol{x};\boldsymbol{D}) = \phi^{\mathrm{s}}_{\boldsymbol{\mathcal{M}}}(f(\boldsymbol{x});\boldsymbol{D}), \tag{2.10}$$

which forms the material distribution on the concerned surface $\boldsymbol{S}$ (as shown in Fig. 4(c)). This treatment is well-posed owing to the bijectivity of the mapping $f$.

#### *2.3.2 TDF construction on a surface of non-zero genus*

When the topology of the concerned surface $\boldsymbol{S}$ is complex (e.g., multi-connected, non-zero genus, etc.), the utilized CCM technique can't be applied directly to obtain the conformal mapping and parametric domain. Taking the classical torus surface as an example (shown in Fig. 5(a)), the surface-cutting operation is adopted to generate an intermediate surface $\boldsymbol{S}^*$ (a simply-connected open surface with zero genus). The TDF construction of $\boldsymbol{S}^*$ is defined as $\phi^{\mathrm{s}}_{\boldsymbol{S}^*}(\boldsymbol{x};\boldsymbol{D}) = \phi^{\mathrm{s}}_{\boldsymbol{\mathcal{M}}}(f^*(\boldsymbol{x});\boldsymbol{D})$ where $f^*\colon \boldsymbol{S}^* \to \boldsymbol{\mathcal{M}}$ is the conformal mapping obtained by the CCM technique (shown in Fig. 5(b)). Considering that the bijectivity is destroyed after the cutting process and $\phi^{\mathrm{s}}_{\boldsymbol{S}^*}(\boldsymbol{x})$ takes different values on different sides of a specific cutting line $\Gamma_i$ on $\boldsymbol{S}$ (i.e., $\Gamma_i'$ and $\Gamma_i''$ on $\boldsymbol{S}^*$, see Fig.

5(a) for reference), the TDF of the original surface $\phi_{\boldsymbol{S}}^{\mathrm{s}}(\boldsymbol{x};\boldsymbol{D})$ is defined as:

$$\phi_{\boldsymbol{S}}^{\mathrm{s}}(\boldsymbol{x};\boldsymbol{D})=\begin{cases}\phi_{\boldsymbol{S}^*}^{\mathrm{s}}(f^*(\boldsymbol{x});\boldsymbol{D}), & \text{if } \boldsymbol{x}\in \boldsymbol{S}\backslash\Gamma_i, \quad (2.11a)\\ \max\Big(\phi_{\boldsymbol{S}^*}^{\mathrm{s}}(f^*(\boldsymbol{x}');\boldsymbol{D}),\phi_{\boldsymbol{S}^*}^{\mathrm{s}}(f^*(\boldsymbol{x}'');\boldsymbol{D})\Big), & \text{if } \boldsymbol{x}\in\Gamma_i, \quad (2.11b)\end{cases}$$

where $f^*(\boldsymbol{x}')$ and $f^*(\boldsymbol{x}'')$ denote the values of $f^*$ taking on $\boldsymbol{x}\in\Gamma_i'$ and $\boldsymbol{x}\in\Gamma_i''$, respectively.

### *2.3.3 TDF definition via the multi-patch stitching scheme*

In principle, the aforementioned procedures can be applied to any surface that has manifold property. Nevertheless, the constructed global parameterizations are normally too stiff, which will cause some numerical instabilities during the construction of $\phi_{\boldsymbol{S}}^{\mathrm{s}}(\boldsymbol{x};\boldsymbol{D})$. This issue is solved by the multi-patch stitching scheme. In this scheme, the surface $\boldsymbol{S}$ is decomposed into $N_{\mathcal{U}}$ patches, that is, $\boldsymbol{S}=\bigcup_{k=1}^{N_{\mathcal{U}}}\boldsymbol{\mathcal{U}}_k$, and the vector of design variables is assembled as $\{\boldsymbol{D}^k\}_{k=1,2,\dots,N_{\mathcal{U}}}$, where $\boldsymbol{D}^k=\left(\boldsymbol{D}_1^k,\boldsymbol{D}_2^k,\dots,\boldsymbol{D}_{n_k}^k\right)$ represents the component set of the $k$-th patch. There's no constraint of the specific shape of patch $\boldsymbol{\mathcal{U}}_k(k=1,2,\dots,N_{\mathcal{U}})$, and the TDF $\phi_{\boldsymbol{\mathcal{U}}_k}^{\mathrm{s}}$ can be determined through the procedures described in section 2.3.1 and 2.3.2.

Taking an eight-shaped torus as the example, the process of the TDF construction based on the multi-patch stitching scheme is depicted in Fig. 6. To begin with, the original surface is decomposed into individual surface patches (shown in Fig. 6(a)). For each patch $\boldsymbol{\mathcal{U}}_k$, the surface-cutting operation is utilized to generate the corresponding intermediate surface $\boldsymbol{\mathcal{U}}_k^*$, and the CCM technique is conducted to obtain the conformal mapping $f_k^*:\boldsymbol{\mathcal{U}}_k^*\to\boldsymbol{\mathcal{M}}_k$, where $\boldsymbol{\mathcal{M}}_k$ denotes the parametric domain. Then, as demonstrated in Fig. 6(b), the TDF $\phi_{\boldsymbol{\mathcal{U}}_k}^{\mathrm{s}}(\boldsymbol{x};\boldsymbol{D}^k)$ of $\boldsymbol{\mathcal{U}}_k$ is defined through Eqs. (2.10-2.11), where $\boldsymbol{D}^k$ represents the design variables of the components placed in parametric domain $\boldsymbol{\mathcal{M}}_k$. For the purpose of constructing the global TDF of $\boldsymbol{S}$, the definition of TDF $\phi_{\boldsymbol{\mathcal{U}}_k}^{\mathrm{s}}(\boldsymbol{x})$ is extended to the whole surface as:

$$\phi_{\boldsymbol{\mathcal{U}}_k^S}^{\mathrm{s}}(\boldsymbol{x};\boldsymbol{D}^k)=\begin{cases}\phi_{\boldsymbol{\mathcal{U}}_k}^{\mathrm{s}}(\boldsymbol{x};\boldsymbol{D}^k), & \text{if } \boldsymbol{x}\in\boldsymbol{\mathcal{U}}_k, \quad (2.12a)\\ \text{const}, & \text{if } \boldsymbol{x}\notin\boldsymbol{\mathcal{U}}_k, \quad (2.12b)\end{cases}$$

where const is a negative number (e.g., $\text{const}=-1$, and the actual value does not affect the optimization result), which represents no material is distributed out of the area $\boldsymbol{\mathcal{U}}_k$ (as illustrated in Fig. 6(c)). In this manner, the global TDF of surface $\boldsymbol{S}$ (as depicted in Fig. 6(d)) can be obtained easily by:

$$\phi_{\boldsymbol{S}}^{\mathrm{s}}(\boldsymbol{x};\boldsymbol{D}) = \max\left(\phi_{\boldsymbol{\mathcal{U}}_1^{\mathrm{s}}}^{\mathrm{s}}, \phi_{\boldsymbol{\mathcal{U}}_2^{\mathrm{s}}}^{\mathrm{s}}, \dots, \phi_{\boldsymbol{\mathcal{U}}_{N_{\mathcal{U}}}^{\mathrm{s}}}^{\mathrm{s}}\right), \tag{2.13}$$

additionally, it's noteworthy that the intersection between two patches is normally non-empty (i.e., $\boldsymbol{\mathcal{U}}_k \cap \boldsymbol{\mathcal{U}}_l \neq \emptyset$), which is necessary for smoothing connections of the components located on neighboring domains. The multi-patch stitching scheme significantly reduces the distortions of parameterization and enhances the fidelity of the description of components, especially for surfaces of large local curvature, high genus, and non-manifold properties.

### *2.4 The description of the embedded components*

Suppose $\boldsymbol{\mathcal{B}}$ is the domain occupied by the concerned thin-walled structure, and $\boldsymbol{S}$ is the middle surface of $\boldsymbol{\mathcal{B}}$. The solid domain $\boldsymbol{\mathcal{B}}$ can be decomposed in the form of Cartesian product, i.e., $\boldsymbol{\mathcal{B}} = \boldsymbol{S} \times \boldsymbol{\omega}$, where $\boldsymbol{\omega} = \left[-\frac{t}{2}, \frac{t}{2}\right]$ representing the variation range of thickness coordinate $\omega$, and the parameter $t$ represents the thickness varying along the surface, and we have $\boldsymbol{S} = \boldsymbol{\mathcal{B}}|_{\omega=0}$. In this manner, the complexity of the original solid domain is transferred to the middle surface, whose TDF can be constructed according to the abovementioned sections. To construct embedded components, the projection operation is resorted to define material distribution. For each point $\boldsymbol{x}$ in domain $\boldsymbol{\mathcal{B}}$, the TDF is defined as

$$\phi_{\boldsymbol{\mathcal{B}}}^{\mathrm{s}}(\boldsymbol{x};\boldsymbol{D}) = \phi_{\boldsymbol{S}}^{\mathrm{s}}(\boldsymbol{x}_{\perp};\boldsymbol{D}), \tag{2.14}$$

where $\boldsymbol{x}_{\perp}$ is the projection point of $\boldsymbol{x}$ to surface $\boldsymbol{S}$, which is obtained by solving the extreme value problem that minimizes the Euclidean distance:

$$\boldsymbol{x}_{\perp} = \mathrm{argmin}_{\boldsymbol{x}_{\boldsymbol{S}} \in \boldsymbol{S}}(\|\boldsymbol{x} - \boldsymbol{x}_{\boldsymbol{S}}\|_2), \tag{2.15}$$

and the thickness coordinate of $\boldsymbol{x}$ (i.e., the distance from point $\boldsymbol{x}$ to its projection point $\boldsymbol{x}_{\perp}$) is defined as $\omega(\boldsymbol{x}) = \|\boldsymbol{x} - \boldsymbol{x}_{\perp}\|_2$. By utilizing the projection operation, the material distribution of domain $\boldsymbol{\mathcal{B}}$ is constructed as $\rho_{\boldsymbol{\mathcal{B}}}^{\mathrm{s}}(\boldsymbol{x};\boldsymbol{D}) = H(\phi_{\boldsymbol{S}}^{\mathrm{s}}(\boldsymbol{x}_{\perp};\boldsymbol{D}))$, where $H(\cdot)$ represents the Heaviside function employed for regularization. To develop a unified framework that can be used to optimize various types of thin-walled structures (e.g., topology optimization, reinforced structure design, sandwich structure design), the following material distribution definitions are proposed:

$$\rho_{\mathcal{B}}^{\mathrm{s}}(\boldsymbol{x};\boldsymbol{D}) = \begin{cases} \rho_1, & \text{if } \omega(\boldsymbol{x}) \in [-\frac{t}{2}, -\frac{t}{2}+\bar{\omega}_2), & (2.16a) \\ H(\phi_{\mathcal{S}}^{\mathrm{s}}(\boldsymbol{x}_\perp;\boldsymbol{D})), & \text{if } \omega(\boldsymbol{x}) \in \left[-\frac{t}{2}+\bar{\omega}_2, \frac{t}{2}-\bar{\omega}_1\right], & (2.16b) \\ \rho_2, & \text{if } \omega(\boldsymbol{x}) \in \left(\frac{t}{2}-\bar{\omega}_1, \frac{t}{2}\right], & (2.16c) \end{cases}$$

where $\bar{\omega}_1$ and $\bar{\omega}_2$ represent two prescribed thicknesses of the base panels, and $\rho_1$ and $\rho_2$ represent two prescribed densities of the base panels, respectively. In this manner, different design problems are distinguished by the thickness coordinate. As demonstrated in Fig. 7, the unified framework can describe various structural forms by taking different values of parameters $\bar{\omega}_1$ and $\bar{\omega}_2$. For instance, when the values of prescribed thicknesses are taken as $\bar{\omega}_1 = 0$ and $\bar{\omega}_2 = 0$, the TDF describes a structure for topology optimization (as shown in Fig. 7(b)). When one of the prescribed thicknesses is non-zero, the TDF describes a structure for rib-reinforcing design (as shown in Fig. 7(c)). When these two prescribed thicknesses are non-zero, the TDF describes a structure for sandwich structure design (as shown in Fig. 7(d)). In addition, parameters $\rho_1$ and $\rho_2$ are set to describe cases of multiple materials, considering that ribs and skin may be made of different materials.

# 3. Problem formulation and sensitivity analysis

This section first states the problem formulation in the present MMC-based solution framework. Then, the sensitivity analysis considering a general function is further elaborated on.

### *3.1 Problem statement*

Without loss of generality, a typical optimization problem formulation can be written as:

$$\text{Find } \boldsymbol{D} = (\boldsymbol{D}_1, \boldsymbol{D}_2, \ldots, \boldsymbol{D}_n) \qquad (3.1a)$$

$$\text{Minimize } I = I(\boldsymbol{D}), \qquad (3.1b)$$

$\mathrm{S.t.}$

$$g_i(\boldsymbol{D}) \leq 0, i = 1, \dots, n_g \tag{3.1c}$$

$$h_i(\boldsymbol{D}) = 0, i = 1, \dots, n_h \tag{3.1d}$$

$$\boldsymbol{D} \in \mathcal{U}_{\boldsymbol{D}}, \tag{3.1e}$$

where the symbol $I$ denotes a general objective function and $\mathcal{U}_{\boldsymbol{D}}$ is the admissible set that the design variable vector $\boldsymbol{D}$ belongs to. In Eq. (3.1), $g_i(\boldsymbol{D})$ and $h_i(\boldsymbol{D})$ are inequality and equality constraint functions, respectively. To be specific, as presented in Fig. 8, a static compliance minimization problem under the constraint of a prescribed volume fraction $\overline{\mathrm{V}}$ (taken as the upper bound) defined in a thin-walled structure region is considered. Suppose $\boldsymbol{\mathcal{B}}$ is the domain occupied by the concerned thin-walled structure, the considered design problem under the current MMC-based framework can be formulated in the following form:

$$\text{Find } \boldsymbol{D} = (\boldsymbol{D}_1, \boldsymbol{D}_2, \dots, \boldsymbol{D}_n), \boldsymbol{u}(\boldsymbol{x}) \tag{3.2a}$$

$$\text{Minimize} \quad C = \int_{\boldsymbol{\mathcal{B}}} H(\phi_{\boldsymbol{\mathcal{B}}}^{\mathrm{s}}(\boldsymbol{x}; \boldsymbol{D}))\boldsymbol{F} \cdot \boldsymbol{u} \mathrm{dV} + \int_{\partial \boldsymbol{\mathcal{B}}} \boldsymbol{t} \cdot \boldsymbol{u} \mathrm{dS}, \tag{3.2b}$$

$$\mathrm{S.t.}$$

$$\int_{\boldsymbol{\mathcal{B}}} \mathbb{E}(H(\phi_{\boldsymbol{\mathcal{B}}}^{\mathrm{s}}(\boldsymbol{x}; \boldsymbol{D})) : \boldsymbol{\varepsilon}(\boldsymbol{u}) : \boldsymbol{\varepsilon}(\boldsymbol{v}) \mathrm{dV} = \int_{\boldsymbol{\mathcal{B}}} H\big(\phi_{\boldsymbol{\mathcal{B}}}^{\mathrm{s}}(\boldsymbol{x}; \boldsymbol{D})\big)\boldsymbol{F} \cdot \boldsymbol{v} \mathrm{dV}$$

$$+ \int_{\partial \boldsymbol{\mathcal{B}}} \boldsymbol{t} \cdot \boldsymbol{v} \mathrm{dS}, \forall \boldsymbol{v} \in U_{ad} \tag{3.2c}$$

$$\int_{\boldsymbol{\mathcal{B}}} H(\phi_{\boldsymbol{\mathcal{B}}}^{\mathrm{s}}(\boldsymbol{x}; \boldsymbol{D})) \mathrm{dV} \leq \int_{\boldsymbol{\mathcal{B}}} \overline{\mathrm{V}} \mathrm{dV}, \tag{3.2d}$$

$$\boldsymbol{u} \in U_{\boldsymbol{u}}, \boldsymbol{D} \in \mathcal{U}_{\boldsymbol{D}}, \tag{3.2e}$$

where $\boldsymbol{u}$ is the primary displacement belonging to a prescribed constraint set $U_{\boldsymbol{u}}$ and $\boldsymbol{v}$ is the virtual displacement belonging to an admissible set $U_{ad}$. In Eq. (3.2), $H(\cdot)$ is the Heaviside function and $C$ is the static compliance function. The symbols $\boldsymbol{F}$ and $\boldsymbol{t}$ denote the body force density and the prescribed surface traction, respectively. Additionally, $\mathbb{E}(H(\phi_{\boldsymbol{\mathcal{B}}}^{\mathrm{s}}(\boldsymbol{x}; \boldsymbol{D})))$ is the fourth-order elastic tensor of material at the point $\boldsymbol{x}$, which is explicitly controlled by the vector of design variables $\boldsymbol{D}$.

### *3.2 Sensitivity analysis*

Taking a general objective function $I$ into consideration, the variation with respect to the variation of the design variable $d$ has the following form (assuming that the design domain's regularity requirements and some smoothness conditions of $I$ are

satisfied):

$$\delta I = \int_{\boldsymbol{\mathcal{B}}} r(\boldsymbol{u}(\boldsymbol{x}), \boldsymbol{v}(\boldsymbol{x}))\delta\phi_{\boldsymbol{\mathcal{B}}}^{\mathrm{s}}(\boldsymbol{x}; \delta d)\mathrm{dV}, \tag{3.3}$$

where $r(\boldsymbol{u}(\boldsymbol{x}), \boldsymbol{v}(\boldsymbol{x}))$ denotes a functional of $\boldsymbol{u}(\boldsymbol{x})$ and $\boldsymbol{v}(\boldsymbol{x})$, $\boldsymbol{u}(\boldsymbol{x})$ and $\boldsymbol{v}(\boldsymbol{x})$ represent the primary and adjoint displacement fields and satisfy $\boldsymbol{u}(\boldsymbol{x}) = -\boldsymbol{v}(\boldsymbol{x})$ in the concerned compliance minimization problem. The symbol $\delta\phi_{\boldsymbol{\mathcal{B}}}^{\mathrm{s}}(\boldsymbol{x}; \delta d)$ denotes the variation of $\phi_{\boldsymbol{\mathcal{B}}}^{\mathrm{s}} = \phi_{\boldsymbol{\mathcal{B}}}^{\mathrm{s}}(\boldsymbol{x}; \boldsymbol{D})$ with respect to the variation of design variable $d$ (i.e., $\delta d$). Since $\phi_{\boldsymbol{\mathcal{B}}}^{\mathrm{s}}(\boldsymbol{x}; \boldsymbol{D}) = \phi_{\boldsymbol{\mathcal{B}}}^{\mathrm{s}}(\phi_{\boldsymbol{\mathcal{S}}}^{\mathrm{s}}(\boldsymbol{x}_{\perp}; \boldsymbol{D}))$, we have $\delta\phi_{\boldsymbol{\mathcal{B}}}^{\mathrm{s}}(\boldsymbol{x}; \delta d) = \frac{\partial\phi_{\boldsymbol{\mathcal{B}}}^{\mathrm{s}}}{\partial\phi_{\boldsymbol{\mathcal{S}}}^{\mathrm{s}}}\delta\phi_{\boldsymbol{\mathcal{S}}}^{\mathrm{s}}(\boldsymbol{x}_{\perp}; \delta d)$. Note that the middle surface is composed of $N_{\mathcal{U}}$ patches (i.e., $\boldsymbol{\mathcal{S}} = \bigcup_{k=1}^{N_{\mathcal{U}}} \boldsymbol{\mathcal{U}}_k$), we have

$$\delta\phi_{\boldsymbol{\mathcal{S}}}^{\mathrm{s}}(\boldsymbol{x}_{\perp}; \delta d) = \sum_{k=1}^{N_{\mathcal{U}}} \partial\phi_{\boldsymbol{\mathcal{S}}}^{\mathrm{s}}(\boldsymbol{x}_{\perp}; \delta d)/\partial\phi_{\boldsymbol{\mathcal{U}}_k^s}^{\mathrm{s}}\delta\phi_{\boldsymbol{\mathcal{U}}_k^s}^{\mathrm{s}}(\boldsymbol{x}_{\perp}; \delta d), \tag{3.4}$$

with

$$\delta\phi_{\boldsymbol{\mathcal{U}}_k^s}^{\mathrm{s}}(\boldsymbol{x}_{\perp}; \delta d) = \begin{cases} \dfrac{\partial\phi_{\boldsymbol{\mathcal{U}}_k}^{\mathrm{s}}(\boldsymbol{x}_{\perp}; \boldsymbol{D}^k)}{\partial d}\delta d, & \text{if } \boldsymbol{x}_{\perp} \in \boldsymbol{\mathcal{U}}_k, \quad (3.5\text{a}) \\ 0, & \text{if } \boldsymbol{x}_{\perp} \notin \boldsymbol{\mathcal{U}}_k, \quad (3.5\text{b}) \end{cases}$$

since $\phi_{\boldsymbol{\mathcal{U}}_k}^{\mathrm{s}}$ is defined by the TDF $\phi_{\boldsymbol{\mathcal{U}}_k^*}^{\mathrm{s}}$ through Eq. (2.11), we have

$$\frac{\partial\phi_{\boldsymbol{\mathcal{U}}_k}^{\mathrm{s}}(\boldsymbol{x}_{\perp}; \boldsymbol{D}^k)}{\partial d} = \frac{\partial\phi_{\boldsymbol{\mathcal{U}}_k}^{\mathrm{s}}(\boldsymbol{x}_{\perp}; \boldsymbol{D}^k)}{\partial\phi_{\boldsymbol{\mathcal{U}}_k^*}^{\mathrm{s}}(\boldsymbol{x}_{\perp}; \boldsymbol{D}^k)}\frac{\partial\phi_{\boldsymbol{\mathcal{U}}_k^*}^{\mathrm{s}}(\boldsymbol{x}_{\perp}; \boldsymbol{D}^k)}{\partial d}. \tag{3.6}$$

Recalling that the TDF $\phi_{\boldsymbol{\mathcal{U}}_k^*}^{\mathrm{s}}$ is defined by the TDF $\phi_{\boldsymbol{\mathcal{M}}_k}^{\mathrm{s}}$ in parametric domain $\boldsymbol{\mathcal{M}}_k$ through Eq. (2.10), i.e., $\phi_{\boldsymbol{\mathcal{U}}_k^*}^{\mathrm{s}}(\boldsymbol{x}_{\perp}; \boldsymbol{D}^k) = \phi_{\boldsymbol{\mathcal{M}}_k}^{\mathrm{s}}(f_k^*(\boldsymbol{x}_{\perp}); \boldsymbol{D}^k)$, where $f_k^*: \boldsymbol{\mathcal{U}}_k^* \to \boldsymbol{\mathcal{M}}_k$ is the conformal mapping obtained by the CCM technique and the symbol $\boldsymbol{\mathcal{M}}_k$ denotes the parametric domain corresponding to the patch $\boldsymbol{\mathcal{U}}_k^*$, we have

$$\frac{\partial\phi_{\boldsymbol{\mathcal{U}}_k^*}^{\mathrm{s}}(\boldsymbol{x}_{\perp}; \boldsymbol{D}^k)}{\partial d} = \frac{\partial\phi_{\boldsymbol{\mathcal{M}}_k}^{\mathrm{s}}(f_k^*(\boldsymbol{x}_{\perp}); \boldsymbol{D}^k)}{\partial d}. \tag{3.7}$$

In Eq. (3.7), the computational method of $\partial\phi_{\boldsymbol{\mathcal{M}}_k}^{\mathrm{s}}(f_k^*(\boldsymbol{x}_{\perp}); \boldsymbol{D}^k)/\partial d$ is consistent with the MMC approach developed for the flat 2D case [37]. It's worth noting that the numerical implementation of the above sensitivity analysis is achieved based on finite element discretization in the physical domain.

## 4. Solution techniques

Up to now, we have completed the task of constructing embedded components and characterizing the material distribution of the thin-walled structure. However, there is still a gap between the definition mentioned above and practical computation. Specifically, for a given point $\boldsymbol{x}$, it is difficult to determine the nearest point $\boldsymbol{x}_{\perp}$ on the surface $\boldsymbol{S}$ and to obtain the distance $\omega(\boldsymbol{x})$. To address these issues, we propose an offset-based solid mesh generation scheme in this section. Additionally, we provide detailed explanations of the solution to the structural responses and the DOF removal technique.

### *4.1 Offset-based solid mesh generation scheme*

Considering the difficulty in solving Eq. (2.15) to find the projection point and determining the thickness coordinate, this subsection proposes an offset-based solid mesh generation technology to numerically simplify the task of computing TDF $\phi_{\mathcal{B}}^{s}(\boldsymbol{x};\boldsymbol{D})$. Let $\boldsymbol{S}_{\Delta}=\{\boldsymbol{V}_0,\boldsymbol{F}_0\}$ denote the triangulated mesh model of the surface $\boldsymbol{S}$ ($\boldsymbol{S}_{\Delta}$ can be generated from a CAD model or obtained from the result of 3D scanning), where $\boldsymbol{V}_0=\{\boldsymbol{v}_0^l|1\leq l\leq n_v\}$ represents the set of nodes and $\boldsymbol{v}_0^l$ is the Cartesian coordinates of the $l$-th node, and $\boldsymbol{F}_0=\{\boldsymbol{f}_0^m|1\leq m\leq n_f\}$ represents the set of elements and $\boldsymbol{f}_0^m=(f_0^{m1},f_0^{m2},f_0^{m3})$ denotes the vertices' indexes in the $m$-th element. Taking the $m$-th element $\boldsymbol{f}_0^m$ for example (shown in Fig. 9(a)), its unit normal vector $\boldsymbol{N}_{f_0^m}$ is determined as:

$$\boldsymbol{N}_{f_0^m}=\frac{\boldsymbol{L}_0^{m1}\times\boldsymbol{L}_0^{m2}}{||\boldsymbol{L}_0^{m1}\times\boldsymbol{L}_0^{m2}||},\tag{4.1}$$

where $\boldsymbol{L}_0^{m1}=\boldsymbol{v}_0^{f_0^{m2}}-\boldsymbol{v}_0^{f_0^{m1}}$ and $\boldsymbol{L}_0^{m2}=\boldsymbol{v}_0^{f_0^{m3}}-\boldsymbol{v}_0^{f_0^{m2}}$ represent two vectors defined by the edges of the element. The first step to construct the solid mesh $\boldsymbol{\mathcal{B}}_{\Delta}$ is generating nodes. Suppose the number of solid elements generated along the thickness direction is $2n_e$, and the total thickness is $t$. Taking the $l$-th node of the surface mesh as example, solid elements' nodes are generated according to the offset operation (shown in Fig. 9(b)):

$$\boldsymbol{v}_j^l=\boldsymbol{v}_0^l+j*\boldsymbol{N}_v^l*\frac{t}{2n_e},-n_e\leq j\leq n_e,\tag{4.2}$$

where $\boldsymbol{N}_v^l$ is the normal vector at the $l$-th node. In classical differential geometry,

suppose $\boldsymbol{S} = \boldsymbol{S}(u, v)$ is a parameterized differentiable surface, the unit normal vector at a specific point $\boldsymbol{S}(u_0, v_0)$ is defined as $\boldsymbol{N}(u_0, v_0) = \boldsymbol{S}_u(u_0, v_0) \times \boldsymbol{S}_v(u_0, v_0)/\|\boldsymbol{S}_u(u_0, v_0) \times \boldsymbol{S}_v(u_0, v_0)\|$. Nevertheless, the continuity of the surface is destroyed in the discrete mesh. Hence, the so-called 1-ring neighborhood definition is utilized to extend the continuous definition to discrete situations [63]. Denote $\boldsymbol{f}_{0l} = \{\boldsymbol{f}_{0l}^m | 1 \leq m \leq n_F^l\}$ the set of elements that contains node $\boldsymbol{v}_0^l$, the unit normal vector at this node is estimated as (shown in Fig. 9(c)):

$$\boldsymbol{N}_v^l = \frac{\sum_{\boldsymbol{f}_{0l}^m \in \boldsymbol{f}_{0l}} |\boldsymbol{f}_{0l}^m| \boldsymbol{N}_{\boldsymbol{f}_{0l}^m}}{\left\| \sum_{\boldsymbol{f}_{0l}^m \in \boldsymbol{f}_{0l}} |\boldsymbol{f}_{0l}^m| \boldsymbol{N}_{\boldsymbol{f}_{0l}^m} \right\|}, \tag{4.3}$$

where $\boldsymbol{N}_{\boldsymbol{f}_{0l}^m}$ and $|\boldsymbol{f}_{0l}^m|$ are the normal vector and the area of element $\boldsymbol{f}_{0l}^m$, respectively. Looping from $l = 1$ to $n_v$ for every node, the nodes of the solid mesh are generated as demonstrated in Fig. 9(d). In addition, the elements of the solid mesh are also obtained from the offset operation. Looping from $m = 1$ to $n_f$ for every element of the original surface mesh, the $j$-th layer surface element is defined as:

$$\boldsymbol{f}_j^m = \boldsymbol{f}_0^m + jn_v = (f_0^{m1} + jn_v, f_0^{m2} + jn_v, f_0^{m3} + jn_v), \tag{4.4}$$

which means nodes in every layer preserve the topology connection relationships of the original surface mesh (shown in Fig. 9(e)). Further, the $j$-th layer solid element along the positive normal direction is defined as $\boldsymbol{P}_j^m = (\boldsymbol{f}_{j-1}^m, \boldsymbol{f}_j^m)$, which is depicted in Fig. 9(f).

In conclusion, the solid mesh is organized as $\boldsymbol{\mathcal{B}}_\Delta = \{\boldsymbol{V}_{\boldsymbol{\mathcal{B}}}, \boldsymbol{P}_{\boldsymbol{\mathcal{B}}}\}$, where $\boldsymbol{V}_{\boldsymbol{\mathcal{B}}} = \{\boldsymbol{v}_j^l | 1 \leq l \leq n_v, -n_e \leq j \leq n_e\}$ is the node set, and $\boldsymbol{P}_{\boldsymbol{\mathcal{B}}} = \{\boldsymbol{P}_j^m | 1 \leq m \leq n_f, -n_e \leq j \leq n_e\}$ is the element set. Based on the aforementioned definitions of the solid mesh, the barriers to determining the material distribution of the thin-walled structure (i.e., the TDF $\phi_{\boldsymbol{\mathcal{B}}_\Delta}^s(\boldsymbol{v}_j^l; \boldsymbol{D})$), have been removed. More specifically, for each node $\boldsymbol{v}_j^l$, the nearest point to mesh surface $\boldsymbol{S}_\Delta$ is $\boldsymbol{v}_0^l$, which means the problem of solving the projection point has been transferred to the modification of subscripts.

### *4.2 Structural response and discrete sensitivities*

In the current work, structural responses are solved by the finite element method using the C3D6 element (a classical linear displacement solid element defined on the triangular prism) provided in ABAQUS [64]. The ersatz material model [37] is introduced to calculate the stiffness matrixes of elements. Taking the $e$-th element ($1 \le e \le N_e$) for an example, its stiffness matrix can be calculated as:

$$\boldsymbol{k}_e = \int_{\Omega_e} \boldsymbol{B}^\top \boldsymbol{D}_e \boldsymbol{B} \mathrm{dV} = \rho_e \int_{\Omega_e} \boldsymbol{B}^\top \boldsymbol{D}^\mathrm{s} \boldsymbol{B} \mathrm{dV}, \tag{4.5}$$

where the symbol $\Omega_e$ denotes the domain occupied by the $e$-th element, the symbol $\boldsymbol{B}$ represents the strain-displacement matrix, and $\boldsymbol{D}_e = \rho_e \boldsymbol{D}^\mathrm{s}$ with $\rho_e$ and $\boldsymbol{D}^\mathrm{s}$ denoting the ersatz density and the constitutive matrix of the solid material, respectively. In the present study, the ersatz density is interpolated as:

$$\rho_e = \sum\nolimits_{i=1}^{6} H_{\alpha,\varepsilon}\left((\phi_{\mathcal{B}}^\mathrm{s})_{e,i}\right)/6, \tag{4.6}$$

with

$$\mathrm{H}_{\alpha,\epsilon}(x) = \begin{cases} 1, & \text{if } x > \epsilon, \\ \dfrac{3(1-\alpha)}{4}\left(\dfrac{x}{\epsilon} - \dfrac{x^3}{3\epsilon^3}\right) + \dfrac{1+\epsilon}{2}, & \text{if } |x| \le \epsilon, \\ \alpha, & \text{otherwise}, \end{cases} \tag{4.7}$$

representing the regularized Heaviside function ($\epsilon = 0.1$ and $\alpha = 10^{-3}$, respectively, in the present work). In Eq. (4.6), $\left(\phi_{\mathcal{B}_\Delta}^\mathrm{s}\right)_{e,i}$ denotes the value of $\phi_{\mathcal{B}_\Delta}^\mathrm{s}$ taken on the $i$-th node of the $e$-th element. In addition, the abovementioned sensitivity analysis is calculated via the finite element discretization formulation. Specifically, the sensitivity of the compliance function has the following form:

$$\frac{\partial C}{\partial d} = -\boldsymbol{U}^\top \frac{\partial \boldsymbol{K}}{\partial d} \boldsymbol{U} = -\sum_{e}^{N_e} \frac{\partial \rho_e}{\partial d} \boldsymbol{u}_e^\top \boldsymbol{k}_0 \boldsymbol{u}_e, \tag{4.8}$$

with

$$\frac{\partial \rho_e}{\partial d} = \sum\nolimits_{i=1}^{6} \frac{\partial H_{\alpha,\varepsilon}}{\partial (\phi_{\mathcal{B}}^\mathrm{s})_{e,i}} \frac{\partial (\phi_{\mathcal{B}}^\mathrm{s})_{e,i}}{\partial d} /6, \tag{4.9}$$

where $\boldsymbol{U}$ is the displacement field vector of domain $\mathcal{B}_\Delta$, $\boldsymbol{K}$ is the global stiffness matrix, $\boldsymbol{k}_0$ is the stiffness matrix of solid element. In addition, the sensitivity of the solid volume fraction with respect to $d$ is quite straightforward and will not be discussed here.

In addition, the following well-known K-S function [68] is utilized to approximate the aforementioned max operator (i.e., Eq. (2.5), Eq. (2.11), and Eq. (2.13)):

$$\chi = \max(\chi^1, \dots, \chi^n) \approx \ln\left(\sum_{i=1}^{n} \exp\left(l\chi^i\right)\right)/l, \tag{4.10}$$

where $l$ is a relatively large positive number (e.g., $l = 100$).

### *4.3 DOF removal technique*

Owing to the explicit description of structural boundaries under the MMC framework, the material distribution in the present work exhibits clear binary behavior. During the optimization process, we utilize the DOF removal technique [51] to enhance computational efficiency. This technique generates a narrow-band mesh model at each iteration step, as illustrated in Fig. 10 (taking a single loading point as an example). Firstly, a loading test is conducted to ensure that external force and displacement boundary conditions are applied to the structure. Secondly, the loading path identification algorithm [52] is employed to distinguish between components that contribute to the structure and those outside the structure (as shown in Fig. 10(b)). Thirdly, elements and corresponding nodes that are covered by the components contributing to the structure are selected, forming a novel narrow-band mesh (as shown in Fig. 10(d)) used for practical structural analysis. It is noteworthy that the DOF removal technique is similar to the remeshing scheme as both rebuild the mesh model after each iteration. However, the DOF removal technique has a lower cost of generating meshes. In addition, the narrow-band mesh model will not be influenced by grey elements, which improves the accuracy of the solution.

## 5. Optimization procedures and remarks

As a summary of the technical section, the optimization procedures for designing thin-walled structures are concluded as follows. These mainly contain three core steps, i.e., geometry preprocessing, layout of embedded components, and conducting structure optimizations.

To begin with, based on a given mesh surface $\boldsymbol{S}_{\Delta}$ (as an approximation of the middle surface $\boldsymbol{S}$ of the concerned thin-walled structure), the corresponding solid mesh $\boldsymbol{\mathcal{B}}_{\Delta}$ is generated, in which the boundary conditions are applied. The mesh surface is divided into several patches, i.e., $\boldsymbol{\mathcal{S}}_{\Delta} = \bigcup_{k=1}^{N_{\mathcal{U}}} \boldsymbol{\mathcal{U}}_k$. For each patch, the corresponding intermediate surface $\boldsymbol{\mathcal{U}}_k^*$ (a single simply connected open surface with genus zero) is generated through the surface-cutting operation, and the conformal mappings $f_k: \boldsymbol{\mathcal{U}}_k^* \rightarrow \boldsymbol{\mathcal{M}}_k, k = 1$ to $N_{\mathcal{U}}$ are obtained through the CCM technique, where $\boldsymbol{\mathcal{M}}_k$ represents the $k$-th parametric domain.

In the second place, components $\boldsymbol{D}^k$ in each parametric domain $\boldsymbol{\mathcal{M}}_k$ are placed to form the initial TDF $\phi_{\mathcal{M}_k}^{\mathrm{s}}(\boldsymbol{p}; \boldsymbol{D}^k)$. Further, the material distribution of the middle surface $\boldsymbol{\mathcal{S}}$ is characterized by the TDF $\phi_{\mathcal{S}}^{\mathrm{s}}(\boldsymbol{x}; \boldsymbol{D})$ through Eqs. (2.11-2.13). By making use of the projection operation (i.e., Eqs. (2.14-2.16)), the embedded components are laid out in the concerned thin-walled structure.

Once the embedded components are established, the FEA process integrated with the DOF removal technique is invoked to estimate the structural responses and sensitivities, which are further submitted to the optimizer to update design variables. These steps, together with the layout of components, are conducted iteratively until the convergence criterion is satisfied. At last, the final design with clear load-transmission path and structure boundaries is obtained at the cost of fewer design variables and degrees of freedom for FEA. To further explain implementation details, some technical remarks are given below.

*Remark 1* The sensitivity solutions are independent of the Jacobian matrix $\boldsymbol{J}$ between the points in physical space and parametric domain, which significantly simplifies the calculation process.

*Remark 2* The current method utilizes a triangulated surface mesh and a prism solid mesh for simplicity and generality. However, other types of elements, including quadrilateral surface mesh, tetrahedra solid mesh, and high-order elements, can also be incorporated into the framework.

*Remark 3* The embedded components can also be constructed using two outside surfaces instead of the middle surface. This can be achieved by changing the Cartesian product form of the thin-walled structure to $\boldsymbol{\mathcal{B}} = \boldsymbol{\mathcal{S}} \times [0, t]$ or $\boldsymbol{\mathcal{B}} = \boldsymbol{\mathcal{S}} \times [-t, 0]$, and applying the CCM technique to the outside surfaces.

*Remark 4* Another possible approach for constructing embedded components is through solid mapping, which directly maps the thin-walled structure to a three-dimensional parametric domain. However, this approach is challenging due to lacking an analogous mathematical structure of three-dimensional solid domains. [65].

# 6. Numerical examples

This section demonstrates the effectiveness and efficiency of the proposed method through three representative numerical examples with complex geometries. Moreover, it also investigates some key factors governing the design results of thin-walled structures, such as surface curvature, boundary conditions, and volume fraction. Without loss of generality, material properties are set as dimensionless. Specifically, Young's modulus and Poisson's ratio of the isotropic solid material are chosen as E=1 and ν=0.3, respectively. Triangulated prism elements are adopted for finite element analysis in all examples. As the surfaces considered are mainly free-form and difficult to describe by analytical expressions or three-view diagrams, the corresponding models have been made available on the website [66] for interested readers. Unless otherwise stated, the upper bound of the volume fraction for all numerical examples is 40%, and the DOF removal technique is employed. The method of moving asymptotes (MMA) [67] is chosen as the numerical optimizer, and the optimization process terminates when the relative change of each design variable, Tol, between two successive steps falls below a prescribed threshold (i.e., Tol=0.0001).

## *6.1 The cycloid-shape thin-walled plate example*

In this example, we examine the topology optimization problem of a cycloid-shaped thin-walled structure with a thickness of 0.5, as illustrated in Fig. 11(a). The surface mesh contains 69290 elements and 34982 nodes, while the solid mesh is generated using the offset-base solid mesh generation scheme, with nine elements along the thickness direction. The boundary conditions are presented in Fig. 12(a), where the corners are fixed, and a line load of magnitude 200 (the magnitude of the is 100) is applied at the center point along the thickness direction, as seen in Fig. 12(b). Fig. 13 and Fig. 14 depict the mapping process from the middle surface to a standard parametric domain and the initial distribution of 72 components in the thin-walled structure, respectively. Fig. 15 shows the optimization iteration histories, indicating that the structure compliance becomes stable and finally converges to $C^{\text{opt}} = 5.34$ at about the 160th step, after oscillations in previous steps. Additionally, Fig. 15 presents some intermediate optimization results, illustrating that some components gradually form the main load-bearing structure between the loading point and the fixed ends, while others fade away as the iteration proceeds. Finally, Fig. 16 displays the final design, with a result similar to topology optimization of membrane structures.

In order to investigate the influence of curvatures on the final designs, the coordinates of the middle surface of the thin-walled structure are adjusted using the following equation:

$$x^3 = \gamma * ((x^1)^2 + (x^2)^2), \tag{4.1}$$

where the parameter $\gamma$ controls the curvature. Fig. 17 presents the parameter settings considered in the current experiment. The solid mesh generation process is omitted, and the boundary conditions are set the same as in Figure 12, i.e., the corners are fixed, and the center point is subject to a line load. Figure 18 shows the optimized results. Result 1 and result 2 differ slightly in curvature. However, this difference leads to an abrupt shift in the in-plane stress state and ultimately results in different load-bearing structures. Result 2 and result 3 have similar material distributions, but the compliance functions differ considerably due to the curvature. Therefore, optimizing the shape and topography of thin-walled structures has the potential to improve structural performance. As the curvature increases from result 3 to result 6, a circular area gradually forms near the loading point, which effectively reduces the stress concentration phenomenon.

In this study, we also compare the optimized results obtained from a shell-based approach [8] with those obtained from the solid-based approach to demonstrate the ability of solid meshes to handle mechanical details in the thickness direction. Firstly, we verify the correlation between the two methods. Fig. 19(a) depicts a shell model with the same geometry as case 2 in Fig. 17(b), where the corners are fixed, and a vertical force with a magnitude of 100 is applied at the central point. The final design of the shell-based approach is shown in Fig. 19(b), which is similar to the result obtained from the solid-based approach (shown in Fig. 18(b)). The relative error of the compliance function is approximately 0.515%. Based on these results, one may conclude that the solid -based method can be replaced by the shell-based method. However, the next experiment refutes this point of view. Fig. 20(a) displays a model with the same geometry and external load conditions as result 2 in Fig. 18(b), while the fixed boundaries are changed to four points at the bottom surface (which is out of the middle surface). The optimized result is presented in Fig. 20(b), where the final design forms a circular outline and bifurcated bars, instead of directly connecting the loading point and the fixed points.

It should be noted that the aforementioned example (as depicted in Fig. 20) cannot be accurately conducted using shell-based methods due to that the shell model is constructed based on the so-called middle-surface modeling hypothesis. This is a common issue faced by many similar shell-based methods. In contrast, the proposed framework utilizes solid meshes, eliminating the aforementioned problems. However, the use of solid meshes may result in a heavy computational burden, which is addressed in this study by employing the DOF removal technique. To demonstrate the effectiveness and accuracy of this technique, we compare the optimized results with and without the technique using the same geometry and boundary conditions as result 2 in Fig. 18(b). The results show that the average time for FEA is reduced from 118.46 seconds per step to 65.26 seconds per step (a decrease of 44.91%) without compromising the optimized structures, as seen in Fig. 21(b).

### *6.2 The bottle-shape thin-walled structure example*

In the second example, a revolving bottle with a bottom is considered, as depicted in Fig. 22. Initially, a solid mesh is generated from the surface mesh, as shown in Fig. 22(a). The solid mesh contains 117040 elements and 58568 nodes, with a thickness of 4 and 6 elements along the thickness direction. Fig. 22(b) illustrates the boundary conditions, where three concentrated forces with the same magnitude ($|\boldsymbol{F}_1| = |\boldsymbol{F}_2| = |\boldsymbol{F}_3| = 100$) and direction are applied on the middle circle of the bottle (depicted in grey), while the top and bottom edges (illustrated in orange) are clamped.

For such a revolving bottle, the CCM technique can be conducted in different ways. The simplest method is to directly map the surface to a planar parametric domain using the single-patch approach. Alternatively, the surface can be divided into two parts: the bottom can be mapped directly to a parametric domain, while the body surface can be mapped to another after performing the surface-cutting operation. The effects of different mapping strategies can be visually explained through component layouts. Fig. 23 demonstrates the process of component layouts via the single-patch mapping approach, where the bottleneck (shown in red) is mapped to the boundary of the rectangle domain (displayed in Fig. 23(b)). The components are laid in the parametric domain (also shown in Fig. 23(b)) and then mapped to the original structure. However, Fig. 23(c) illustrates that even if this single-patch mapping is bijective and preserves topology, the component layouts of the concerned thin-walled structure are severely twisted due to the enormous and inevitable distortions of the global mapping process.

As depicted in Fig. 24, the body surface of the bottle is segmented by a cut along the red line before being mapped to the parametric domain (Fig. 24(b)). The resulting multi-patch-based component layouts are shown in Fig. 24(c), where the components' configurations are identifiable (for illustration, components are not laid in the bottom part). The multi-patch stitching scheme preserves the fidelity of geometry modeling and the structure's load-bearing performance, and meanwhile significantly reduces the distortions of global mappings under the single-patch framework. Therefore, the CCM strategy illustrated in Fig. 24 is utilized in subsequent bottle designs. As shown in Fig. 25, the initial component layout is set to 3×6 cells. The optimization process's iteration histories are plotted in Fig. 26, from which it is evident that the compliance function converges rapidly to a flat value. At the 30-th step, the main load-bearing structure has formed clearly, and the subsequent optimization process focused on adjusting small components to fine-tune the structure's stiffness.

The final design, with compliance of $C^{\mathrm{opt}} = 0.6462$, is presented in Fig. 27. The top view of the design shows that it is composed of three types of hierarchical components, each with a distinct function for resisting the concentrated force. As demonstrated in Fig. 27(a), the load-bearing components (in red) form a platform at the loading point of $\boldsymbol{F}_1$ (marked by the blue cross) to bear the external load, hence these components are categorized as load-bearing components. However, since the loading point is far from the fixed end, the bridging components (in golden) are assembled to this platform to transfer the load to the fixed end, forming the main load-transmission path (instead of a mechanism). In addition to these, the final design includes a series of reinforcing components (in grey, the third category) that enhance the stiffness and stability of the whole structure. These components are primarily distributed between different bridging components, forming several triangular substructures. In Fig. 27(b), the side view of the final structure illustrates the effects of concentrated force $\boldsymbol{F}_2$, which is applied along the tangential direction of the surface. The load-bearing components (in red) form a claw-shaped substructure to grab the transverse bridging components (in golden) and resist the shear strain around the loading point. As can be seen from the above results, the proposed MMC-based framework has multiple advantages, including clear boundaries of the optimized structures, no grey elements involved, and components with specific mechanical meanings, which is critical for helping engineers understand and interpret the final design.

In this example, the impact of various initial component layouts on optimized results is also examined. As depicted in Fig. 28, different component layout settings were used, where all component layouts of the bottom part are set as 2×2 cells, and the component layouts of the body part are set as 3×6 cells, 4×8 cells, 5×10 cells, and 7×14 cells, respectively, from left to right. Fig. 29 presents the final designs of each setting, revealing that the primary load-bearing structures and the compliance function are relatively similar across all the layouts. However, more details, such as reinforcing components, are added to the final design as the component layout is refined. Therefore, we suggest that the current method can use optimization results with a small number of initial components to construct the main force transmission path, and optimization results with a higher number of initial components can be used to achieve more detailed designs.

Moreover, designing rib-reinforced structures is a crucial aspect of thin-walled structure design, and it can be readily achieved in the current framework. In this example, we present the results of optimizing rib reinforcement structures for two settings. As shown in Figure 30, the initial components layout is shown as 4×8 cells, with the thickness of the base plate (in grey) set to 2, i.e., half of the overall thickness. Figures 31(a) and 31(c) depict the design settings for the two rib reinforcement structures, with Young's modulus of the base plate set to 0.2 and 0.7 and the volume fraction of the stiffener set to 0.225 and 0.10, respectively. Figures 31(b) and 32(d) present the optimization results for the two settings. As shown in Figure 31(b), the load-bearing capacity of the base plate is relatively weak, which results in a similar layout of the stiffeners to the results of topology optimization (i.e., Fig. 29(b)). In contrast, the material volume fraction is much smaller in the second setting, resulting in a clear path for the stiffeners, as shown in Figure 31(d).

### *6.3 The tee-branch pipe example*

The final example considers a tee-branch pipe model to demonstrate the applicability of the proposed algorithm to practical engineering structures. Figure 32(a) shows that the middle surface mesh has 107,784 elements and 54,103 nodes. Figure 32(b) displays the solid mesh generated from the middle surface with a thickness of 2 and 8 elements along the thickness direction, along with the external loads and displacement boundary conditions. Since the topology of the tee-branch pipe is complex, and each pipe branch has a different shape, it is difficult to parameterize the surface directly. To address this problem, a multi-patch stitching scheme is used, as shown in Fig. 33. The original surface is partitioned into four patches according to geometric features, and the fourth patch is sewn with a circle domain (in grey) to form a topological cylinder, as shown in Figure 34. For each patch, the surface-cutting operation is used to obtain an intermediate surface, which is then used to obtain the corresponding parametric domain. With these treatments employed, conformal parameterizations are constructed for these four patches, and the global TDF on the tee-branch pipe is constructed as described in Section 2. Figure 35 displays the initial component layout and material distribution for each patch on the thin-walled structure.

Fig. 36 illustrates iteration histories and shows that the optimization process terminates after 160 iteration steps. Due to the advantage of a few design variables under the MMC framework, the main load-transmission path is obtained after the 40th step, resulting in a final design with $C^{\mathrm{opt}}$ =43.38, as presented in Fig. 37. The layout of components on two loaded pipes is similar to the well-known 2D short beam example [36]. Moreover, four different boundary conditions are applied to the tee-branch pipe to investigate their effect on the final design, as shown in Fig. 38(a-d). The corresponding optimized results are demonstrated in Fig. 38(e-h).

Finally, the sandwich-type reinforced structure design is considered in this example, where the same boundary conditions are applied as in Fig. 32(b), and the initial component layout is the same as in Fig. 35. As illustrated in Fig. 39, the thickness of the outer and inner layers (considered as the undesigned domain) is 0.5, while the thickness of the designed layer is 1. Three numerical examples with different parameter settings are examined. In each case, the volume fractions are set as 0.375, 0.45, and 0.55, and Young's modulus of the undesigned layers is set to 0.15, 0.3, and 0.5, respectively. Young's modulus of the designed layer is set to 1 in all cases. The optimization results are displayed in Fig. 40(a)-(c), it is found that the optimized results mainly consist of sheet structures and truss structures, which cannot be modeled using shell models.

## 7. Conclusions and extensions

This article proposes a novel approach based on the MMC method for optimizing complex thin-walled structures, in which the embedded solid components is employed to construct a general description of thin-walled structures. The middle surface of thin-walled structure is mapped to a standard parametric domain using the CCM technique to characterize material distribution. By integrating the surface-cutting operation and the multi-patch stitching scheme, the proposed algorithm can handle thin-walled structures with complex geometries and significantly reduce numerical instabilities caused by the mapping process. Numerical examples demonstrate that this method has attractive properties of requiring few iterations and achieving fast convergence. The proposed approach obtains structures with clear boundaries at the cost of few design variables and degrees of freedom for finite element analysis (FEA).

Compared to existing shell model-based approaches, the proposed method has several advantages. Firstly, the solid model is not limited by the assumptions made by shell models, making the current method applicable to structures of any thickness. Secondly, the proposed method can handle boundary conditions and material distribution along the thickness direction naturally. Finally, solid meshes ensure precise structural responses, which is crucial for practical structural optimization results.

Notably, the current work's primary purpose is to propose a general framework for various design problems of thin-walled structures. Hence specific details of different scenarios are not considered, e.g., how to control the sizes of reinforced ribs on the curved surface and how to incorporate the constitutive relations of composite materials into the designs of sandwich structures. These topics are both of excellent engineering and scientific value but also very challenging. In addition, this method can be extended to other issues such as dynamics, acoustics, and thermal. Corresponding research results will be reported in separate works.

## Acknowledgment

This work is supported by the National Key Research and Development Plan (2022YFB3303000), the National Natural Science Foundation (11821202, 12002077, 12002073), the Liaoning Revitalization Talents Program (XLYC2001003), the Fundamental Research Funds for the Central Universities (DUT21RC(3)076, DUT20RC(3)020), and 111 Project (B14013).

**Figures**

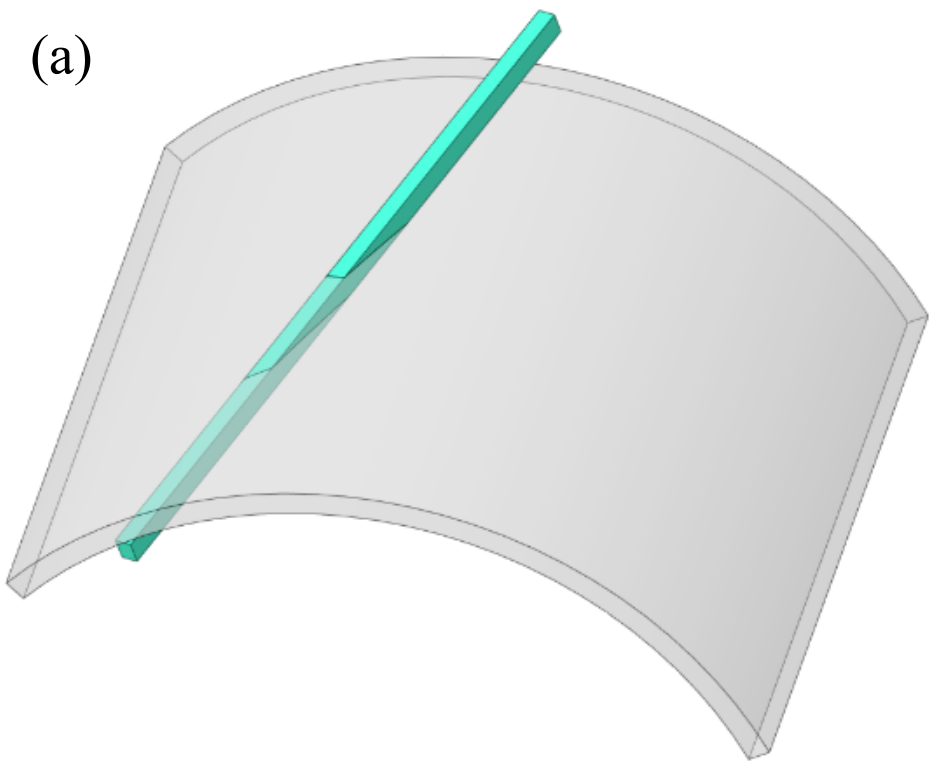

Fig. 1 Schematic illustration of the necessity of the embedded component. (a) The component defined in Euclidean space intrudes into the base panel during the optimization process. (b) The embedded component perfectly follows the shape variation of the thin-walled structure.

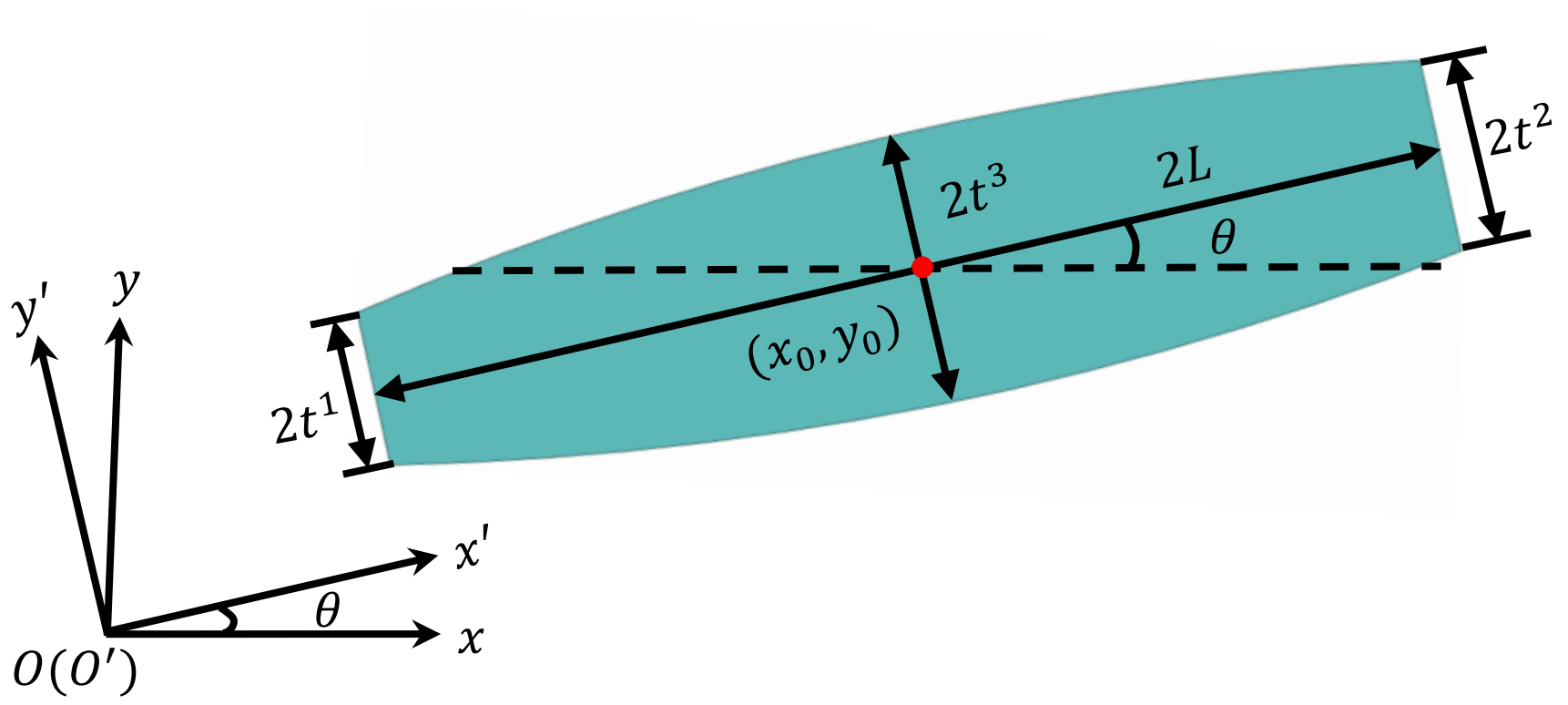

Fig. 2 Schematic illustration of the typical 2D structural component.

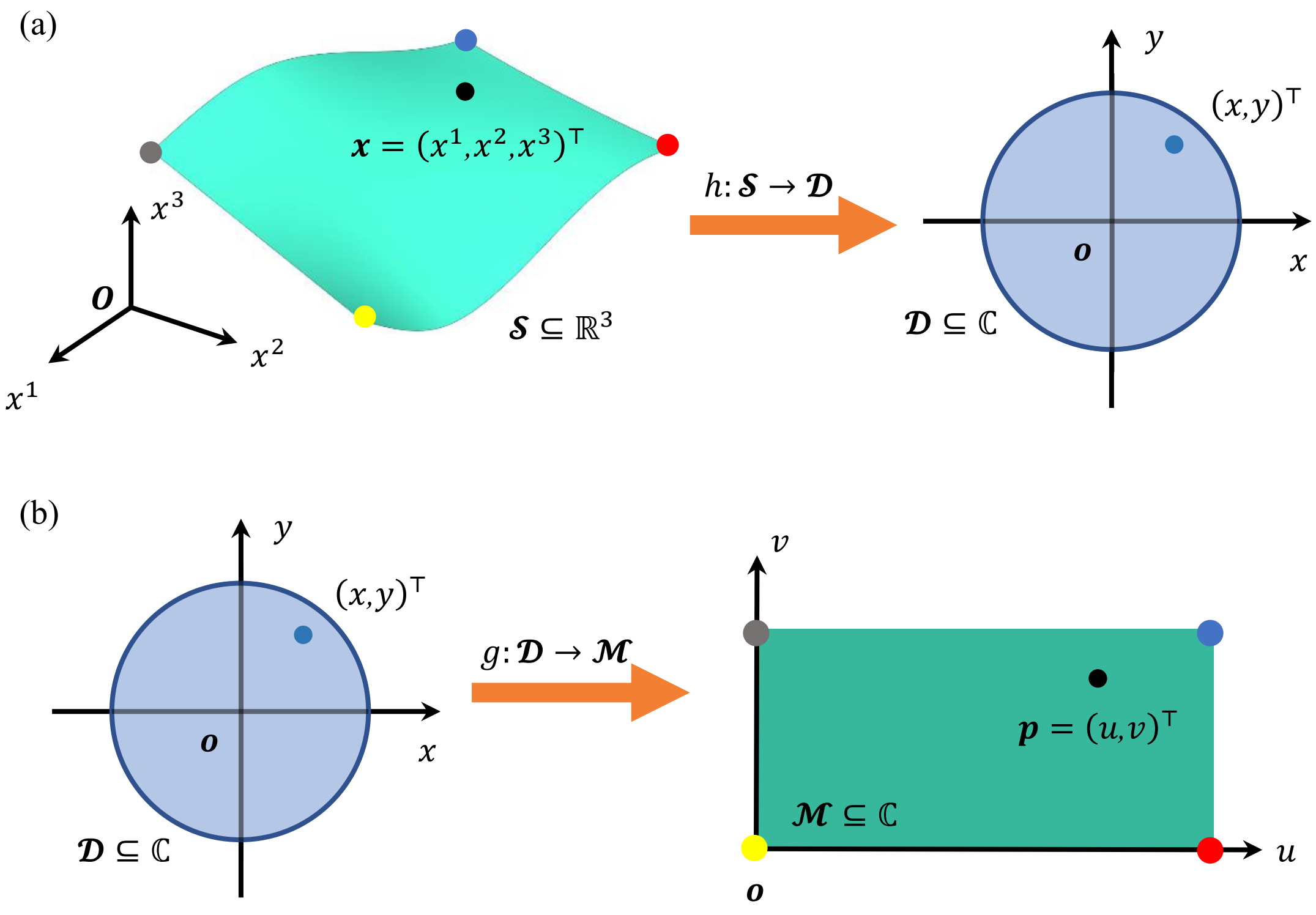

Fig. 3 Schematic illustration of the process of solving the desired conformal mapping by composing two quasi-conformal mappings. (a) The first mapping. (b) The second mapping.

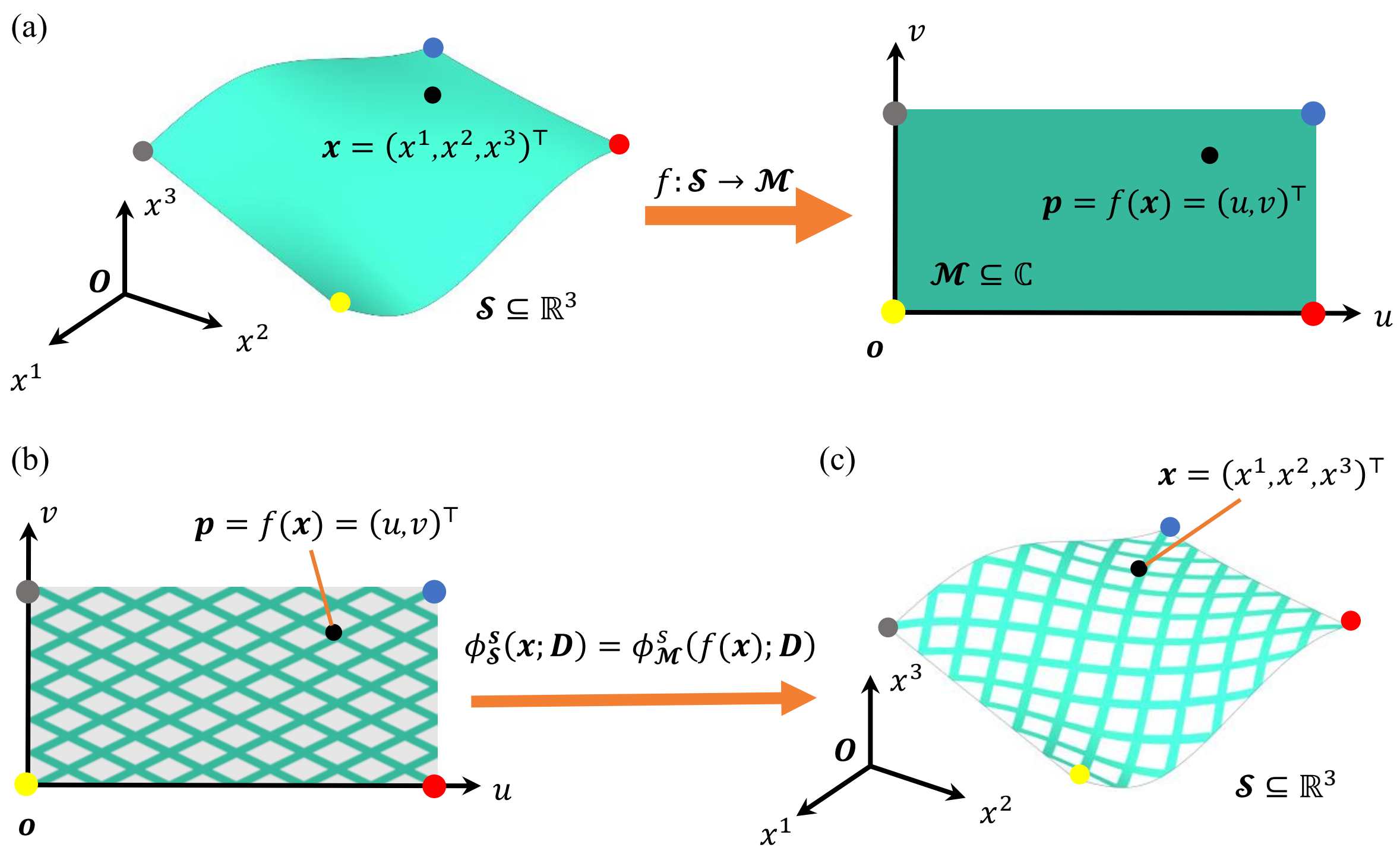

Fig. 4 TDF definition on a simply-connected open surface with genus zero. (a) The conformal mapping and the parametric domain. (b) TDF construction in the parametric domain. (c) TDF construction on the concerned surface.

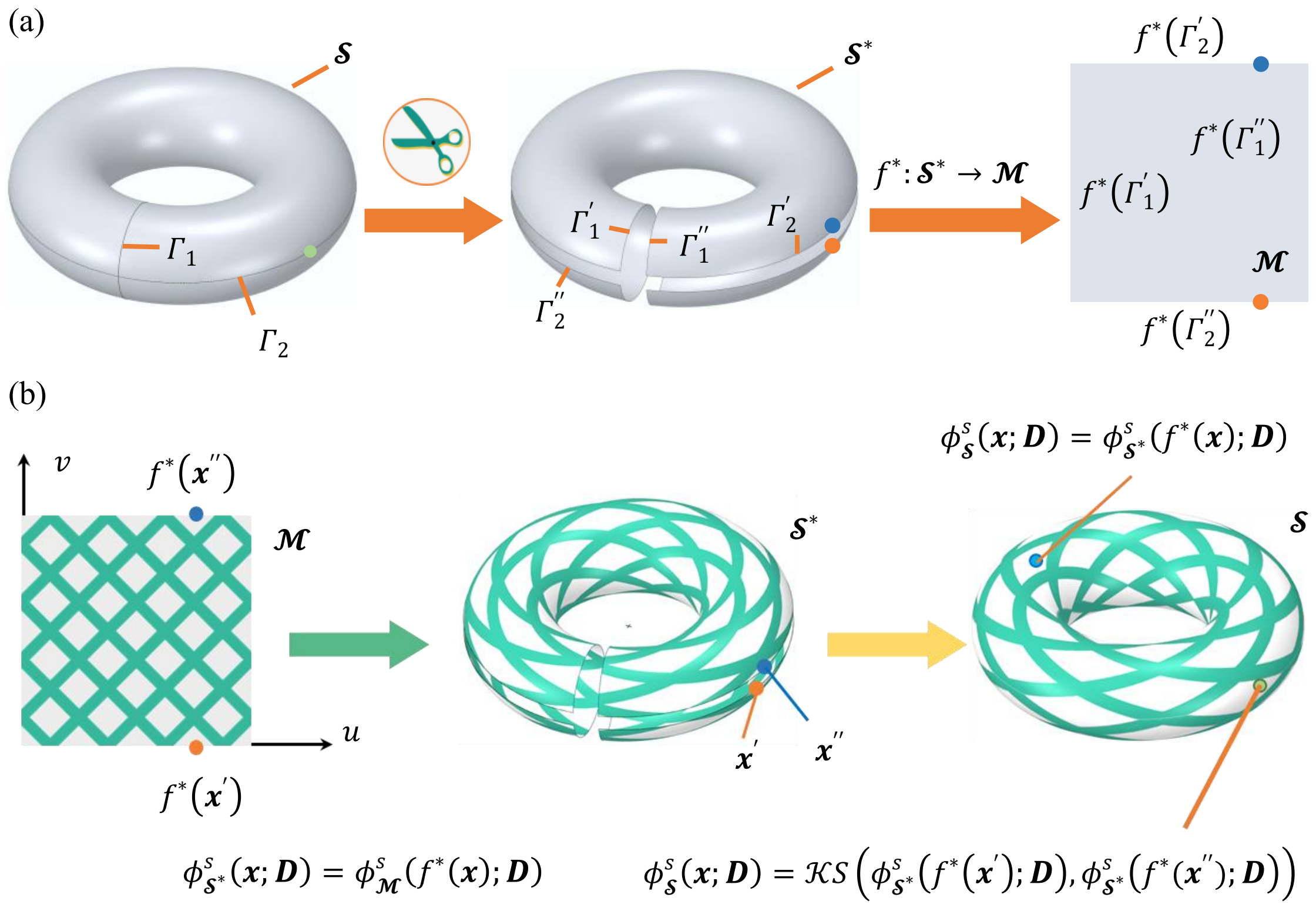

Fig. 5 Parameterization and TDF construction of a surface with non-zero genus. (a) Parameterization of a surface with non-zero genus via the surface-cutting operation. (b) TDF definition on a surface with non-zero genus.

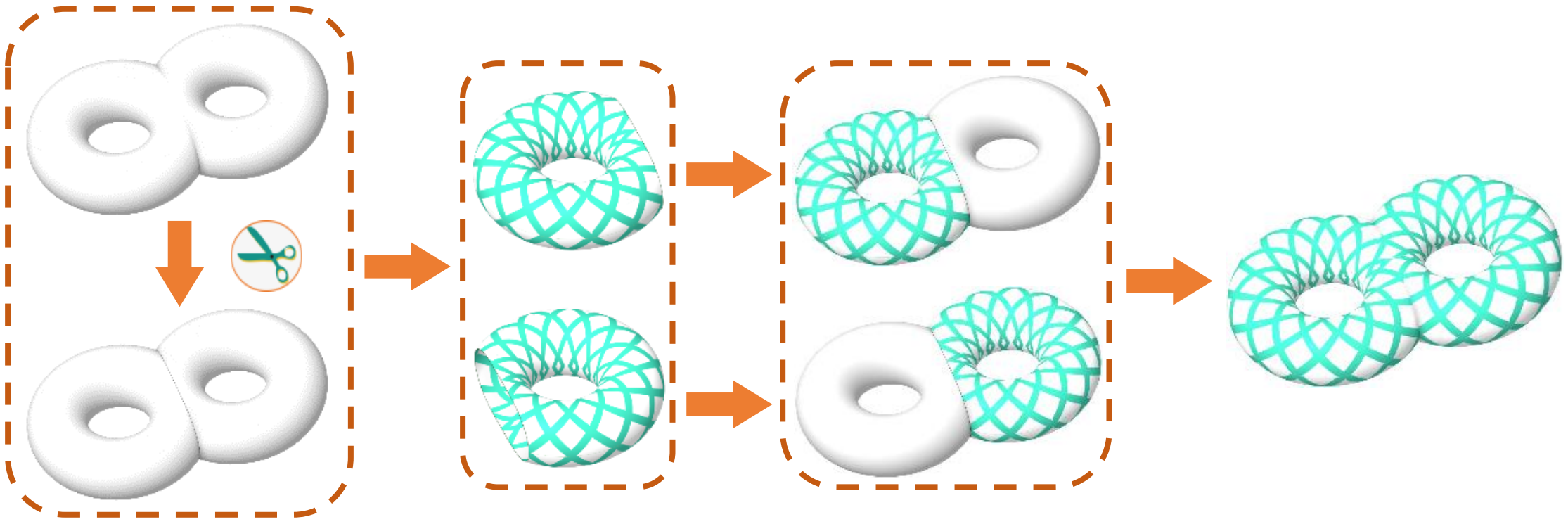

(a) (b) (c) (d)

Fig. 6 TDF construction on a complex surface via the multi-patch stitching scheme. (a) Decomposing the concerned surface into individual patches according to geometric features. (b) TDF construction on each patch. (c) TDF extensions from surface patches to the original surface. (d) TDF definition on the original surface via the multi-patch stitching scheme.

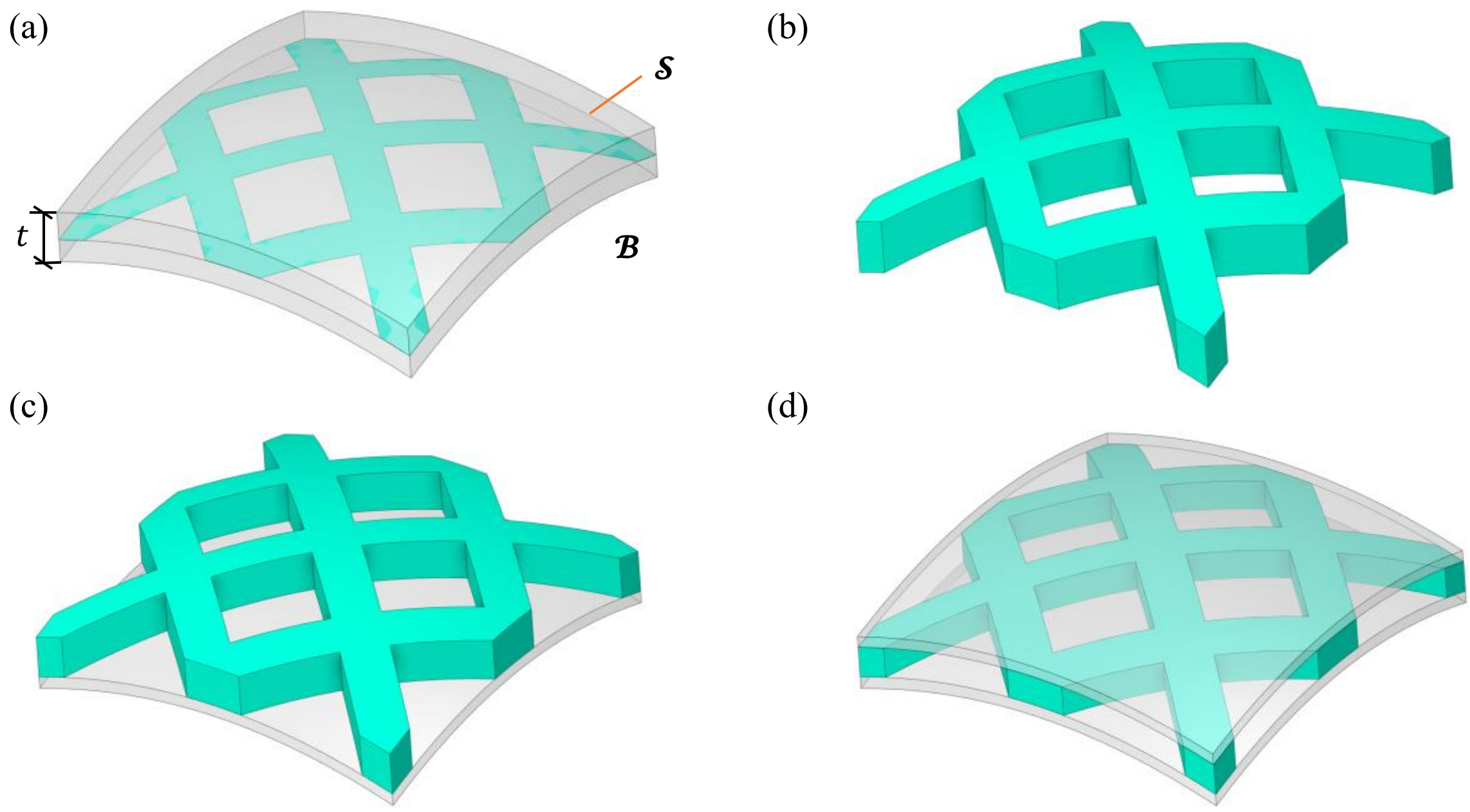

Fig. 7 Schematic illustration of the embedded components. (a) The material distribution of the solid domain is obtained via the projection operation from the middle surface. (b-c) Different application scenarios, i.e., topology optimization, rib-reinforcing structure design, and sandwich structure design, are unified in the proposed framework.

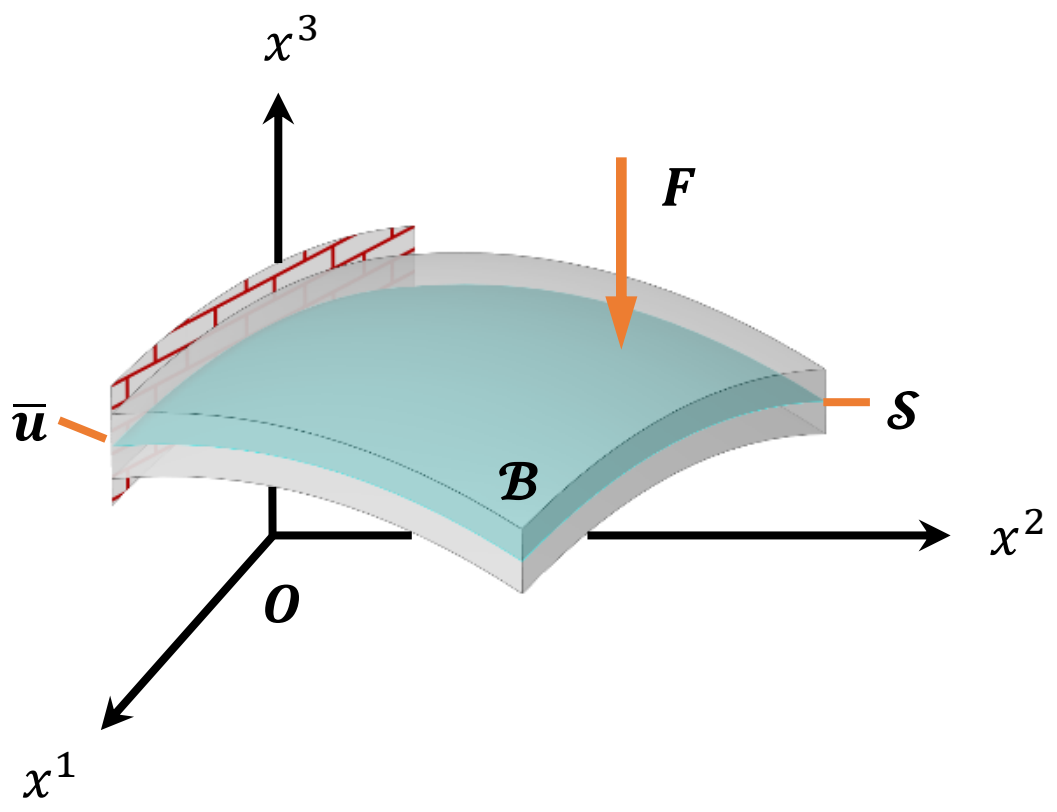

Fig. 8 Minimizing the compliance of a thin-walled structure via topology optimization.

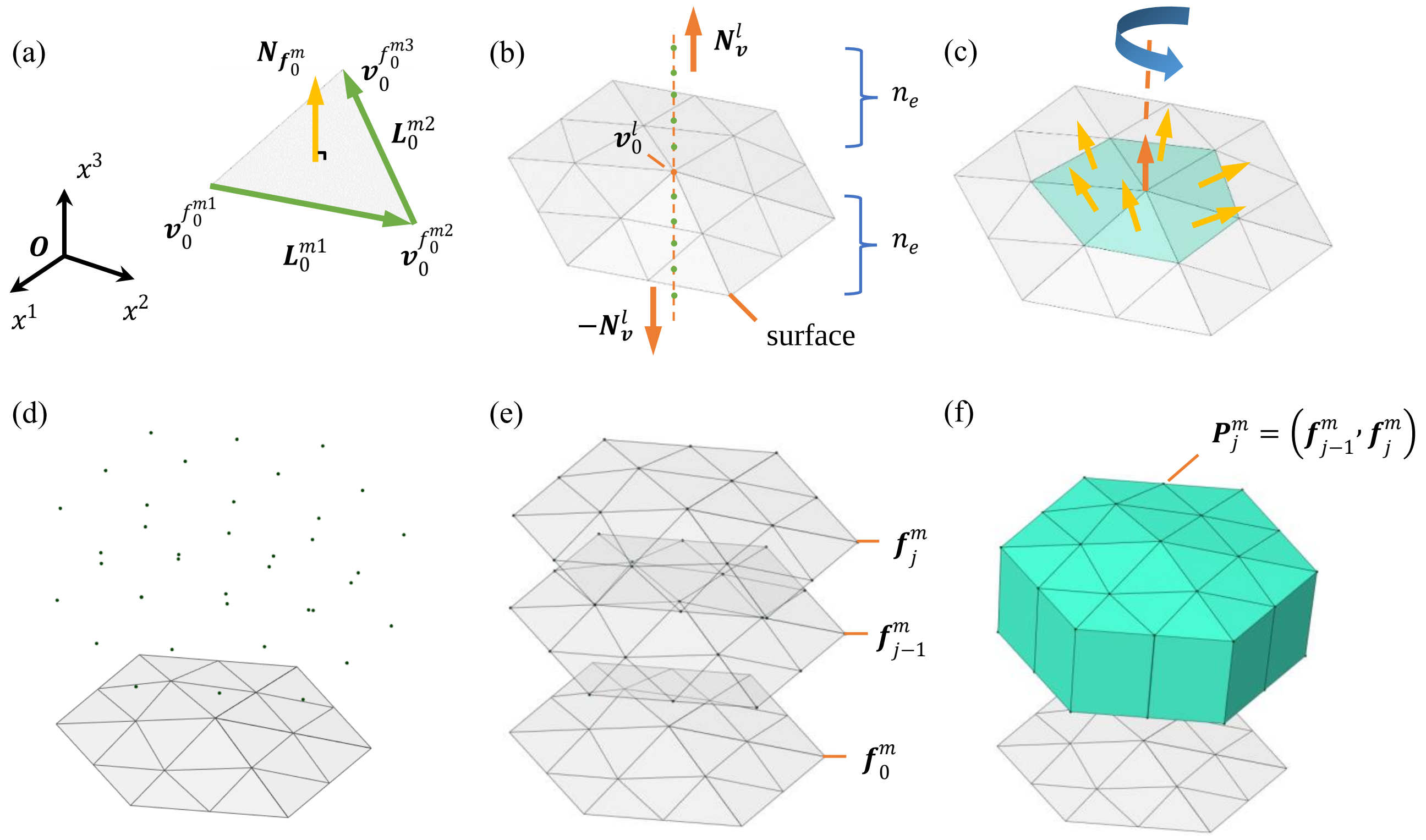

Fig. 9 Schematic illustration of the process of offset-based solid mesh generation scheme. (a) The definition of the unit normal vector of an element. (b) Generating nodes according to the offset operation. (c) Normal vector definition at a node via the 1-ring neighborhood scheme. (d) Generating nodes of the solid mesh (taking the positive normal direction as example). (e) Surface element generation. (f) Solid element generation.

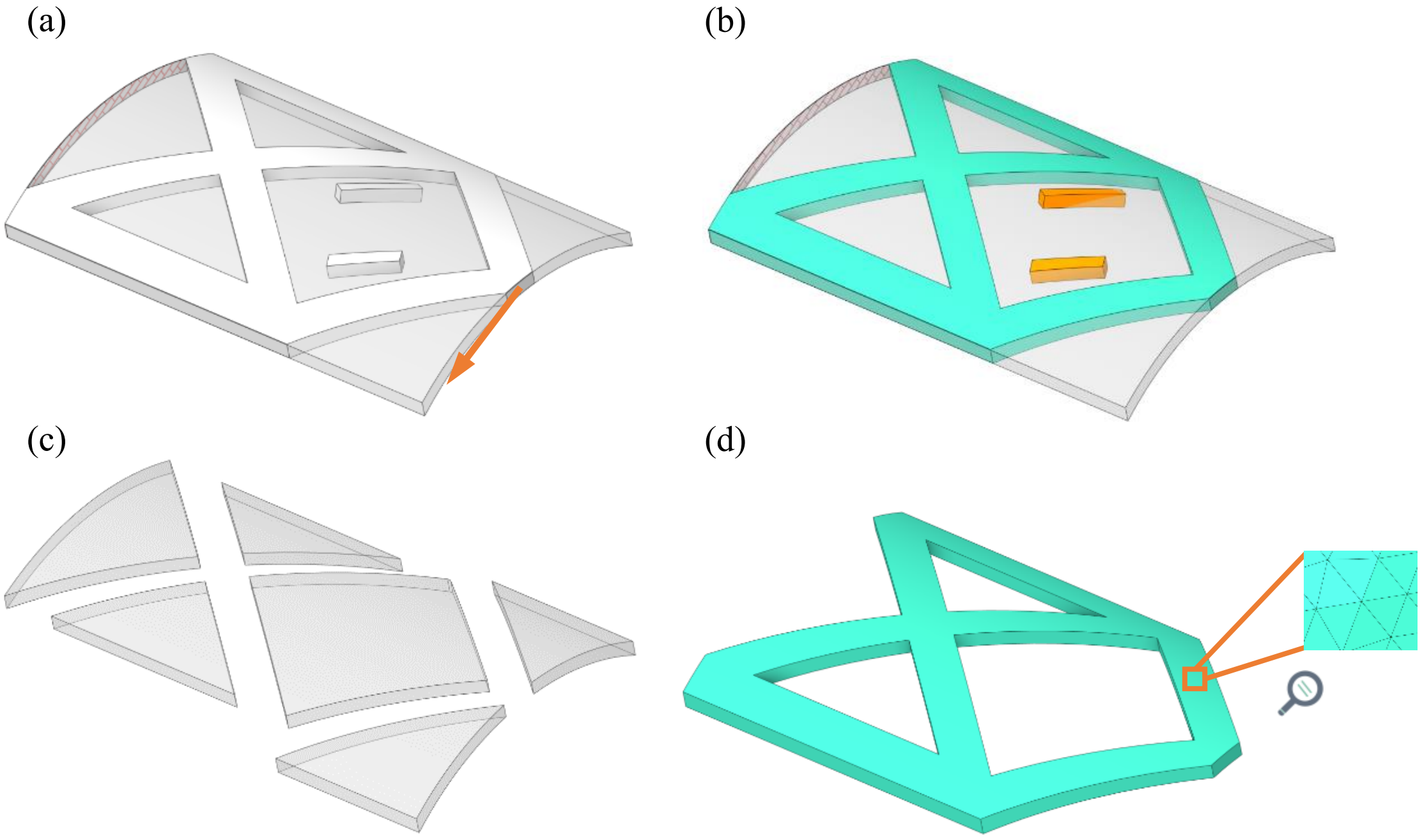

Fig. 10 Schematic diagram of the DOF removal technique. (a) Original computational model. (b) Distinguish structural components (visualized in green) from unstructured components (visualized in orange) according to the loading path identification algorithm. (c) Removed mesh. (d) Narrow-band mesh generated from the DOF removal technique.

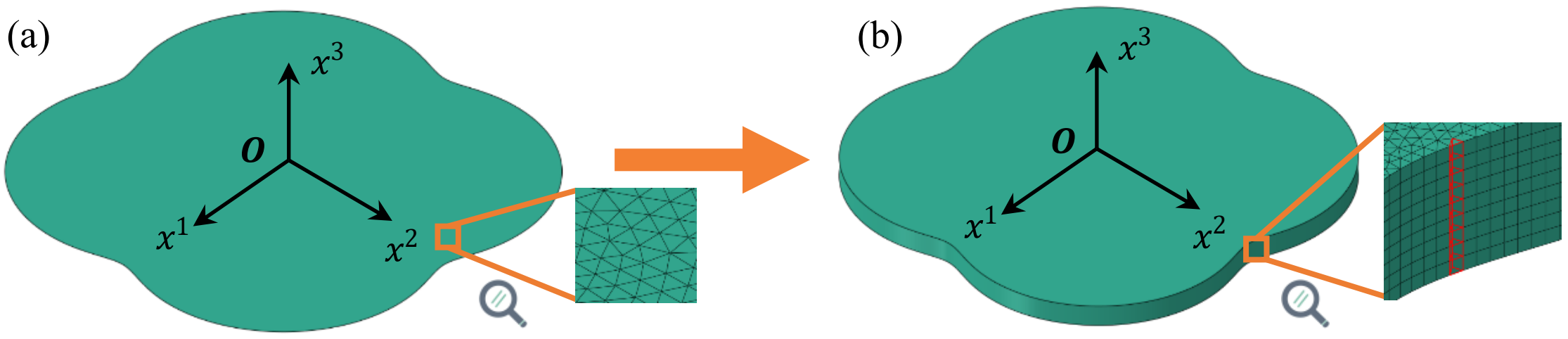

Fig. 11 Generating the solid mesh of the cycloid-shape thin-walled structure. (a) Middle surface mesh. (b) Solid mesh.

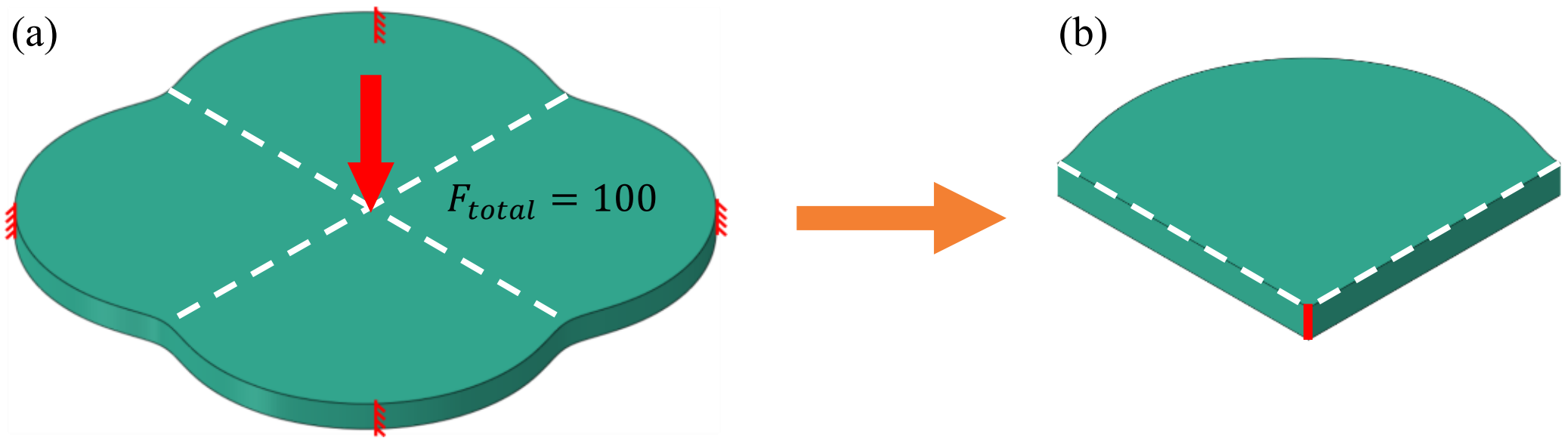

Fig. 12 Boundary conditions of the cycloid-shape thin-walled structure. (a) Displacement and external load boundary conditions. (b) The external load is a line load (visualized in red) applied at the center point along the thickness direction.

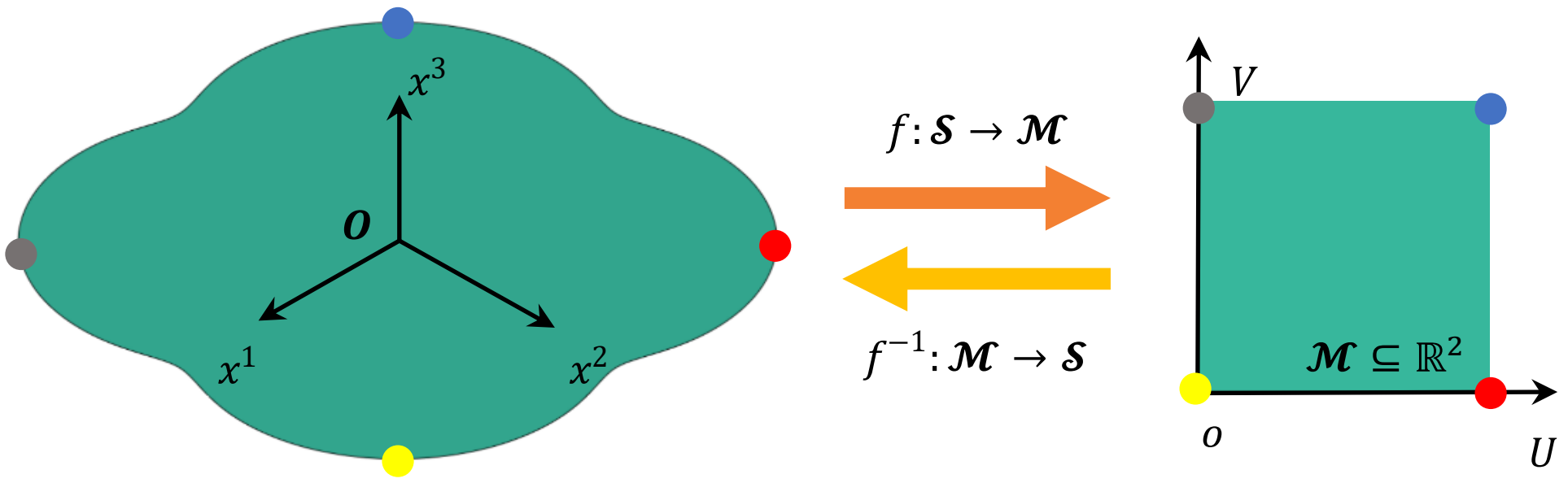

Fig. 13 Conformal mapping of the middle surface of the cycloid-shape thin-walled structure.

(a) (b) (c)

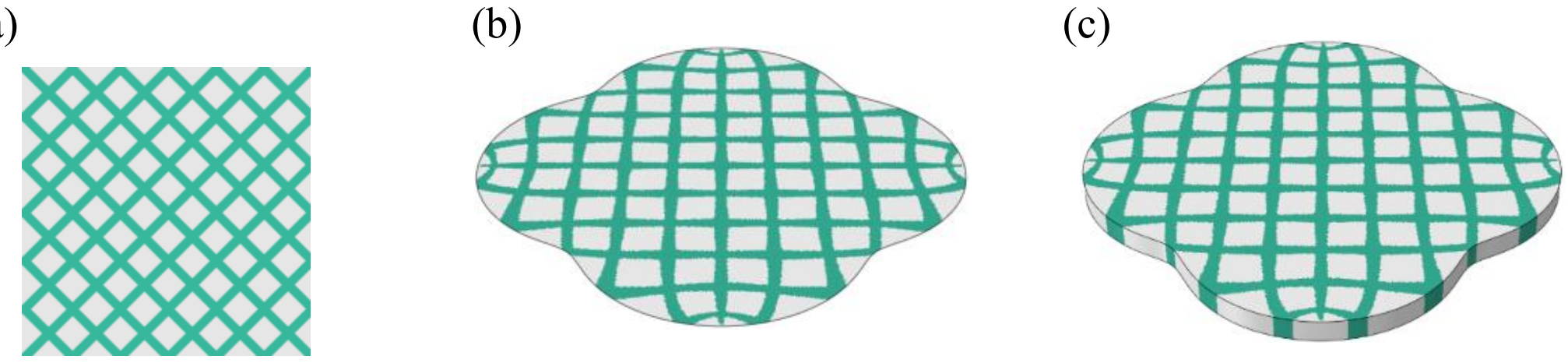

Fig. 14 Initial layout of components in (a) parametric domain; (b) middle surface; (c) solid domain.

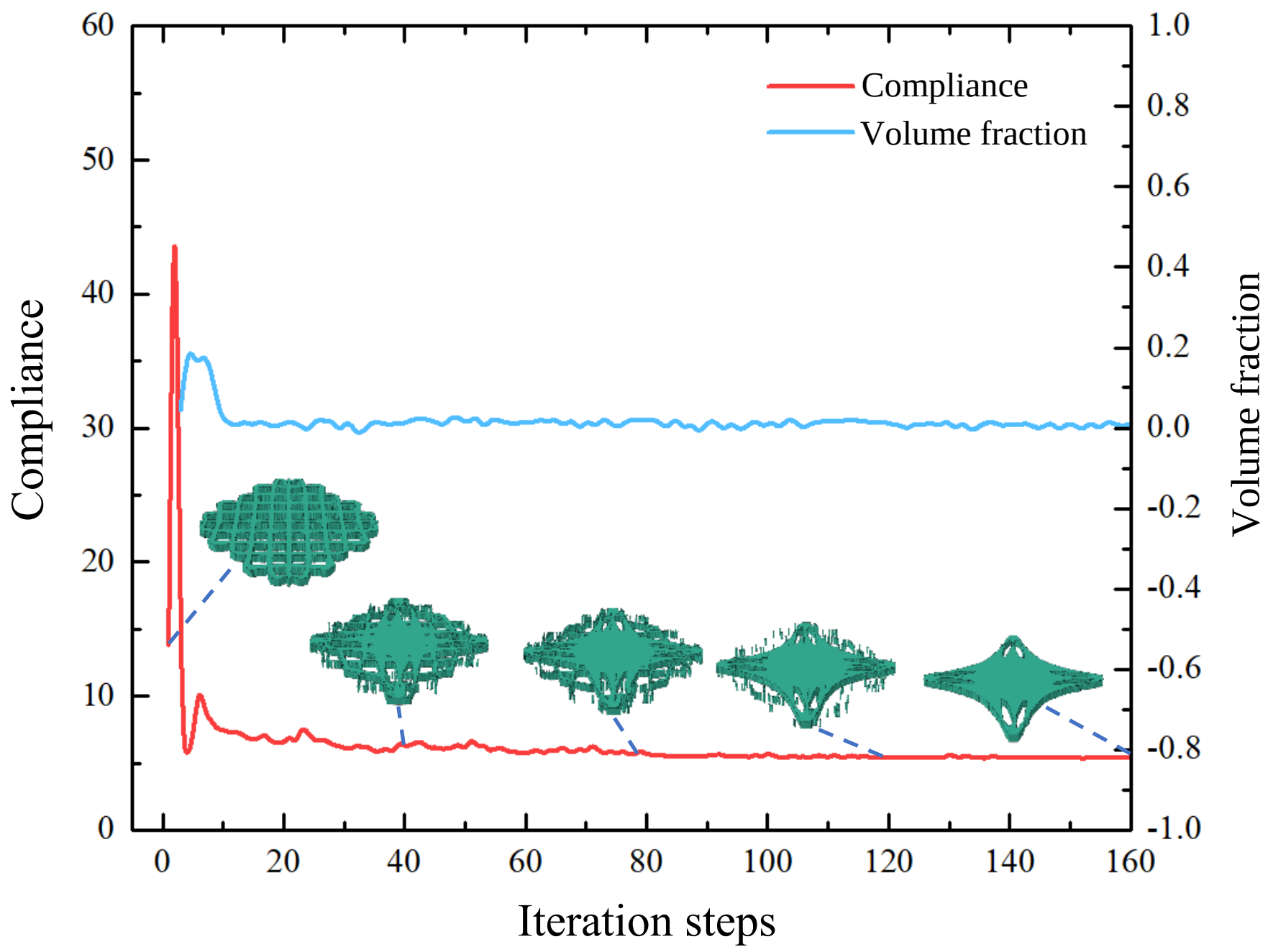

Fig. 15 Iteration histories and intermediate results of the cycloid-shape thin-walled structure.

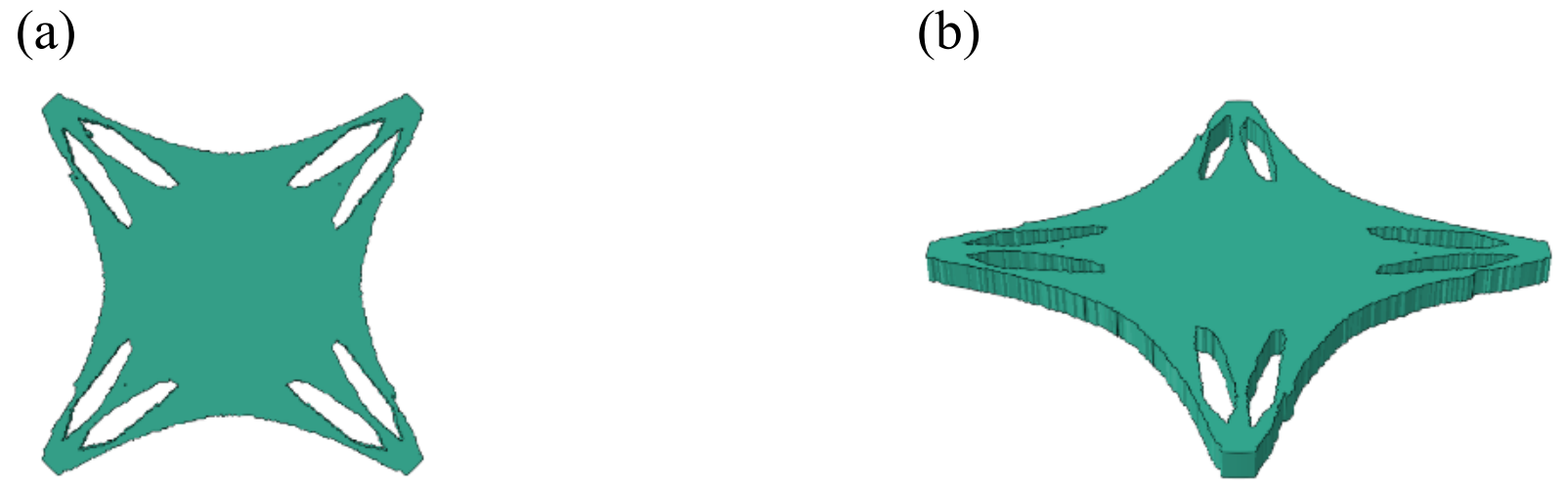

Fig. 16 Optimized design of the cycloid-shape thin-walled structure ($C^{\mathrm{opt}} = 5.34$). (a) Top view. (b) Side view.

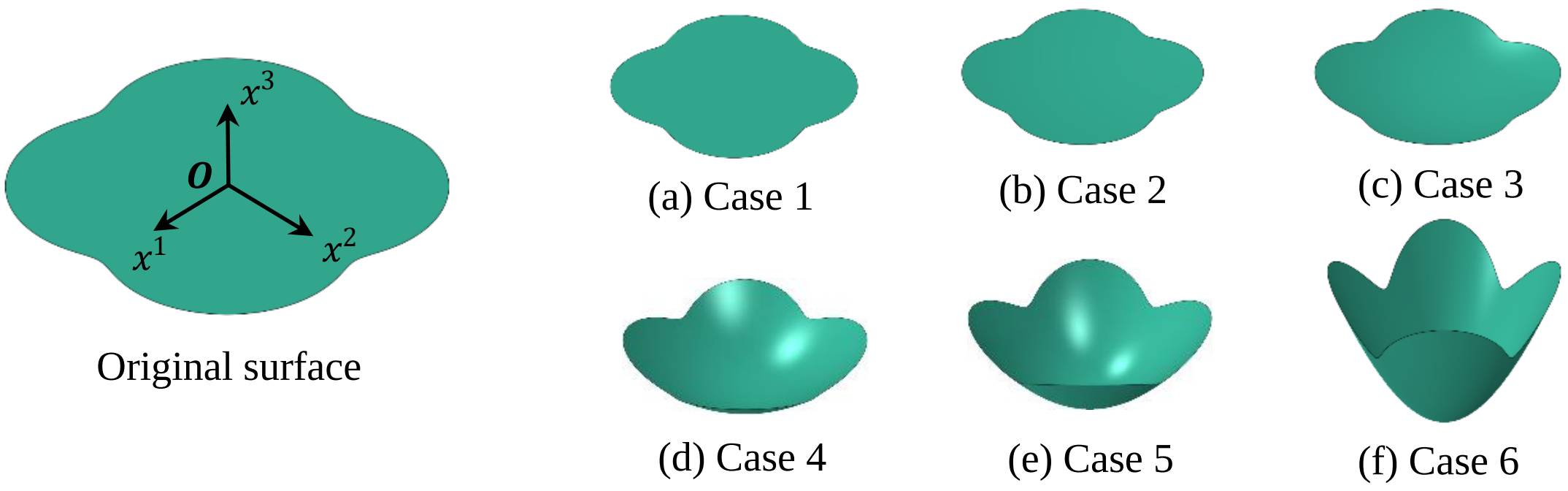

Fig. 17 Illustration of modifications of the middle surface. From case 1 to 6, the parameter $\gamma$ is set as 0, 0.025, 0.05, 0.10, 0.15, 0.30, respectively.

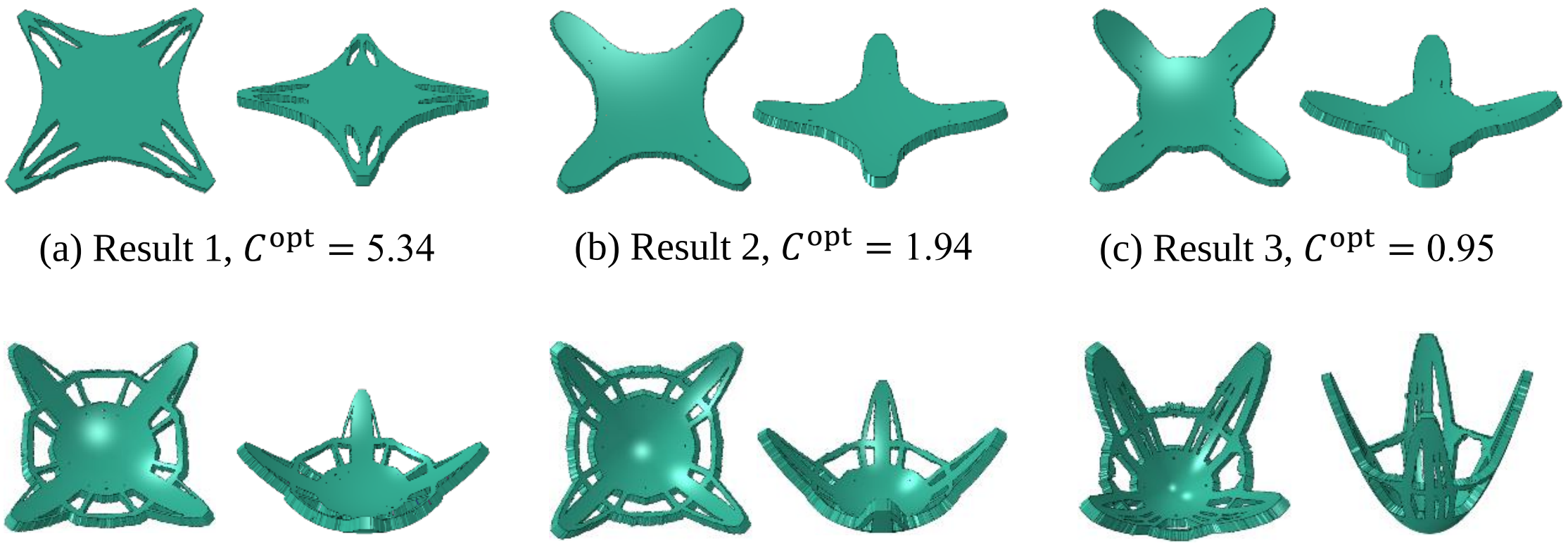

(a) Result 1, $C^{\mathrm{opt}} = 5.34$ (b) Result 2, $C^{\mathrm{opt}} = 1.94$ (c) Result 3, $C^{\mathrm{opt}} = 0.95$

(d) Result 4, $C^{\mathrm{opt}} = 0.54$ (e) Result 5, $C^{\mathrm{opt}} = 0.41$ (f) Result 6, $C^{\mathrm{opt}} = 0.29$

Fig. 18 Comparison of final designs of different curvatures.

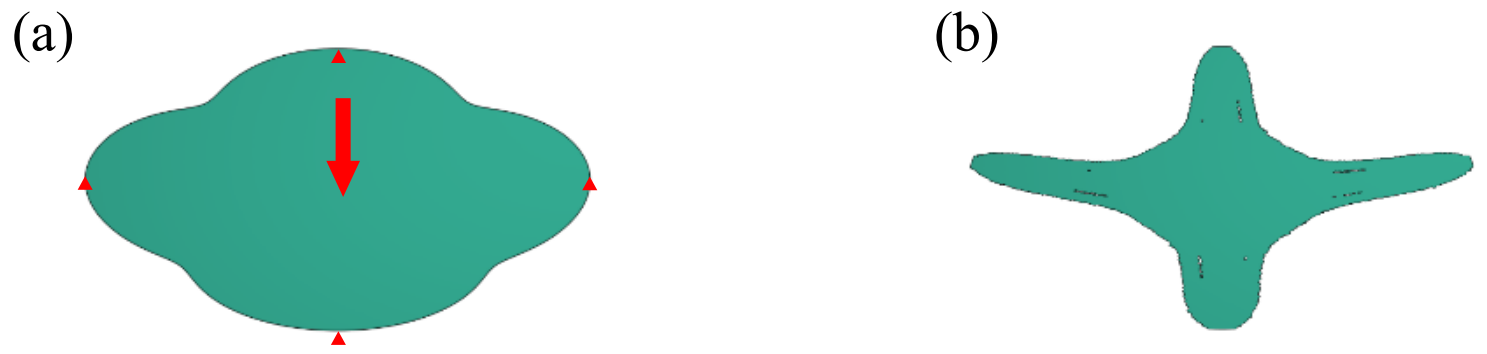

Fig. 19 Optimized result of the shell-based approach. (a) Boundary conditions. (b) Final design ($C^{\mathrm{opt}} = 1.93$).

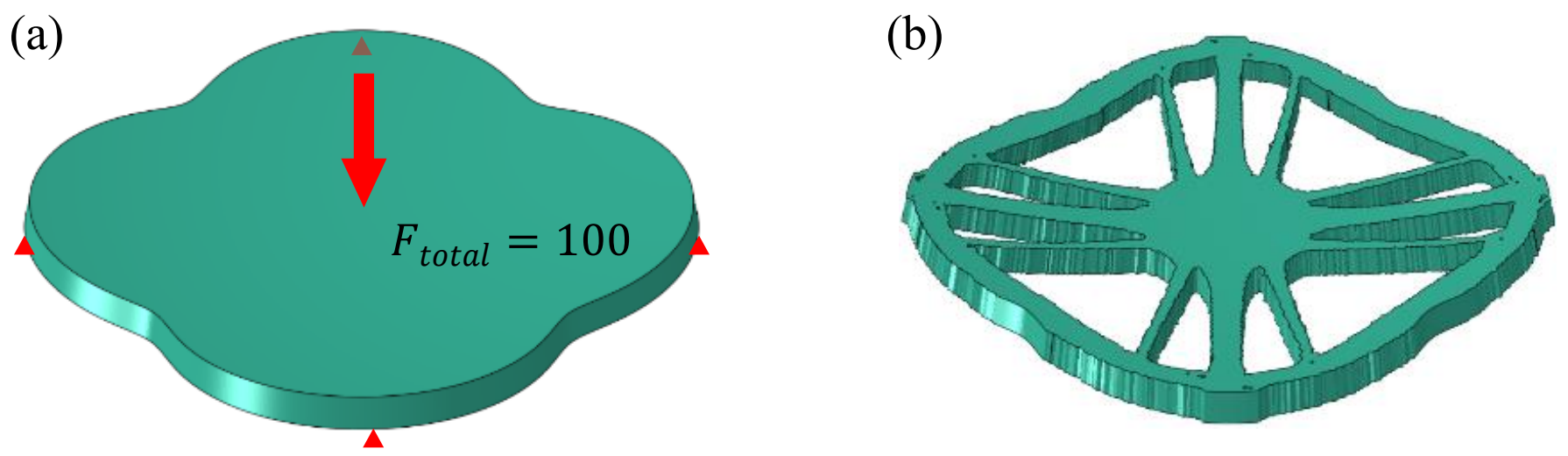

Fig. 20 Structure optimization case with boundary conditions applied out of the middle surface. (a) Boundary conditions. (b) Final design ($C^{\mathrm{opt}} = 8.12$).

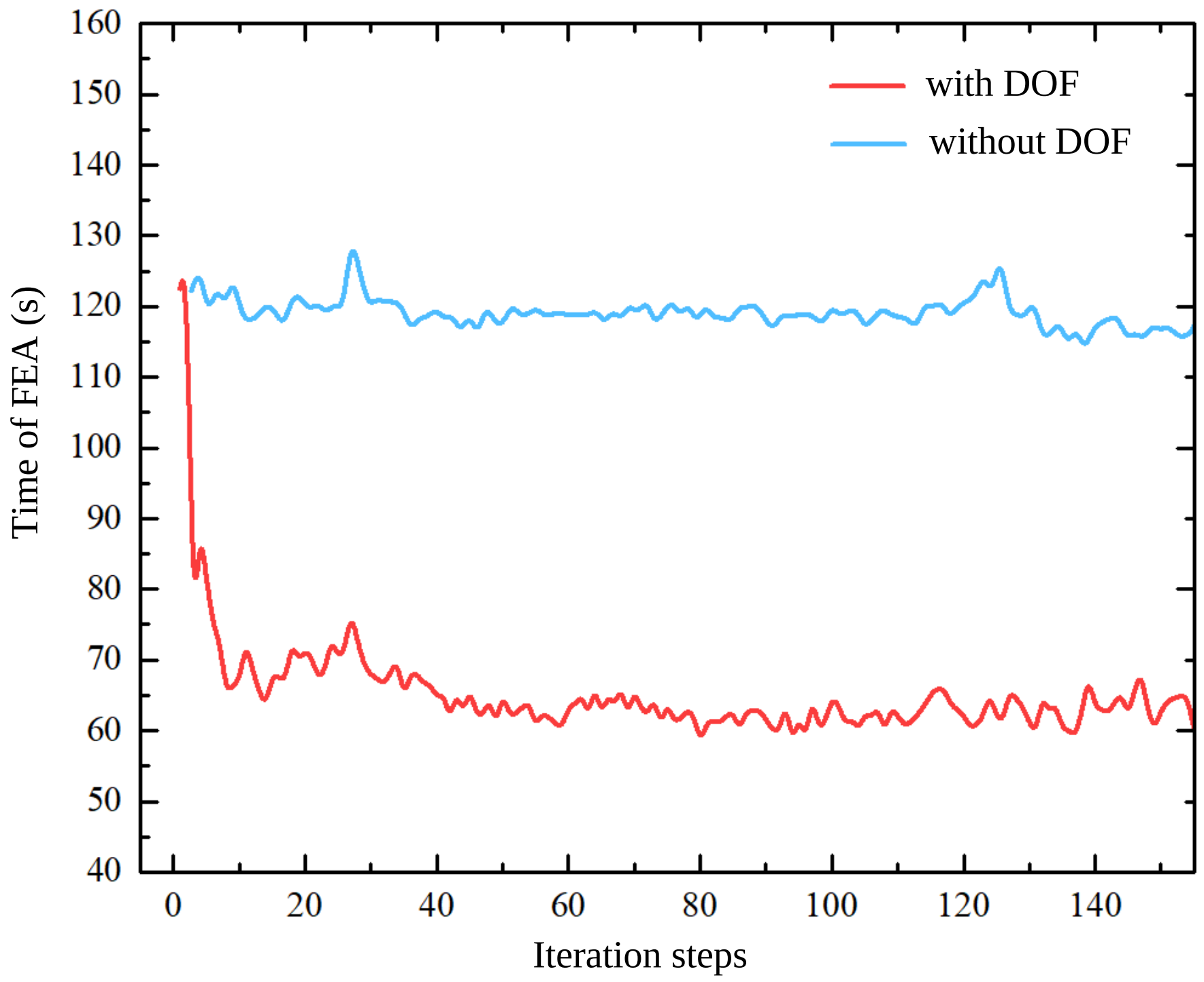

(a) Comparison of FEA times with and without the DOF removal technique.

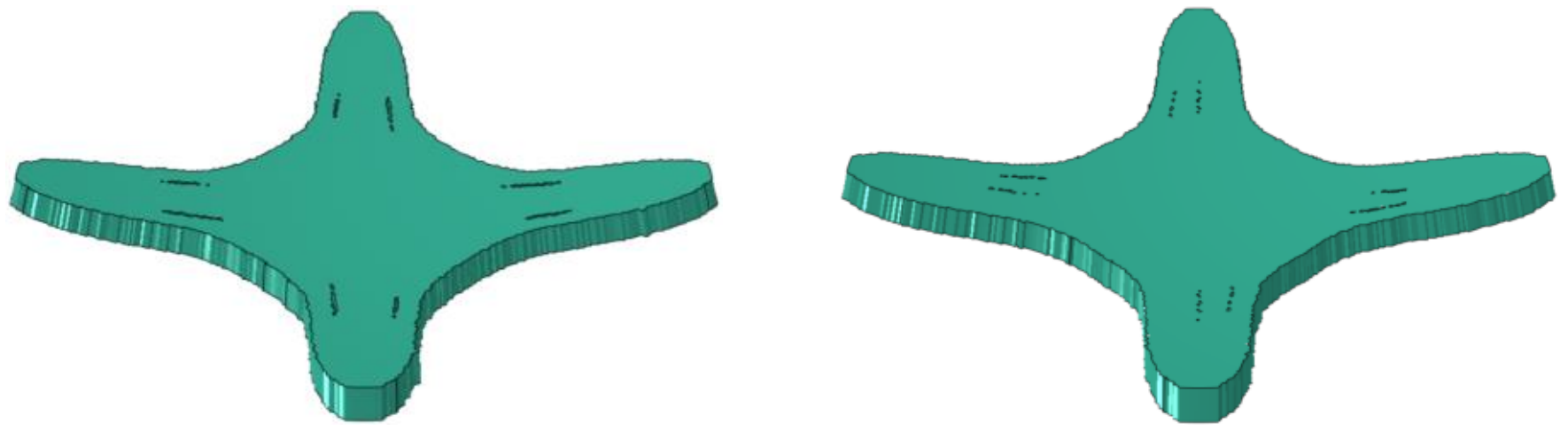

(b) Comparison of final designs with and without the DOF removal technique.

Fig. 21 Optimized results with and without the DOF removal technique.

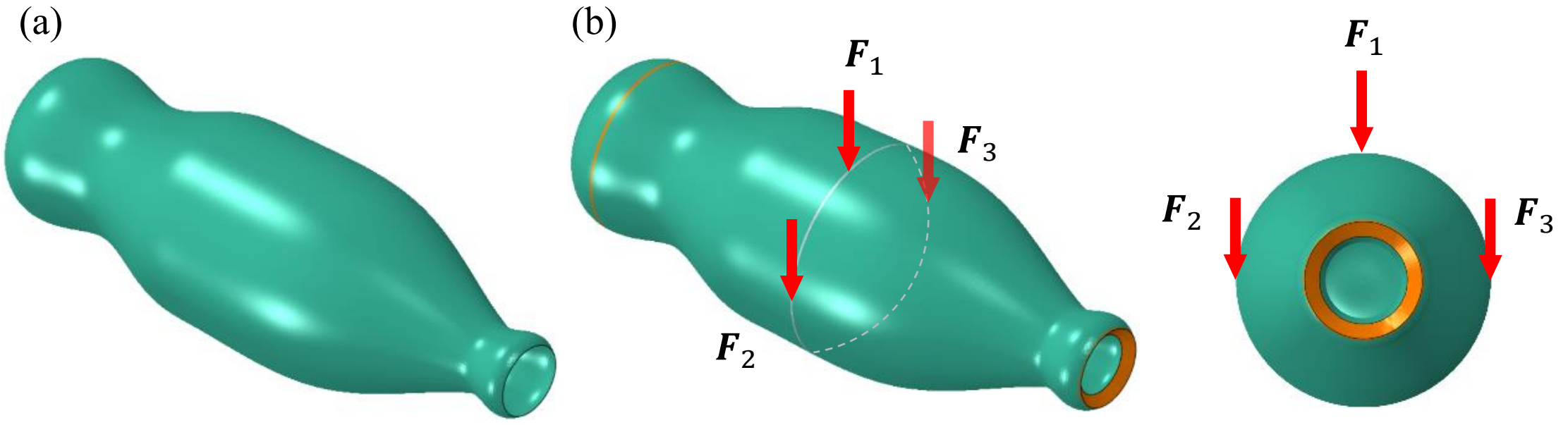

Fig. 22 Geometry and boundary conditions of the bottle example. (a) Middle surface model. (b) Solid model and boundary conditions.

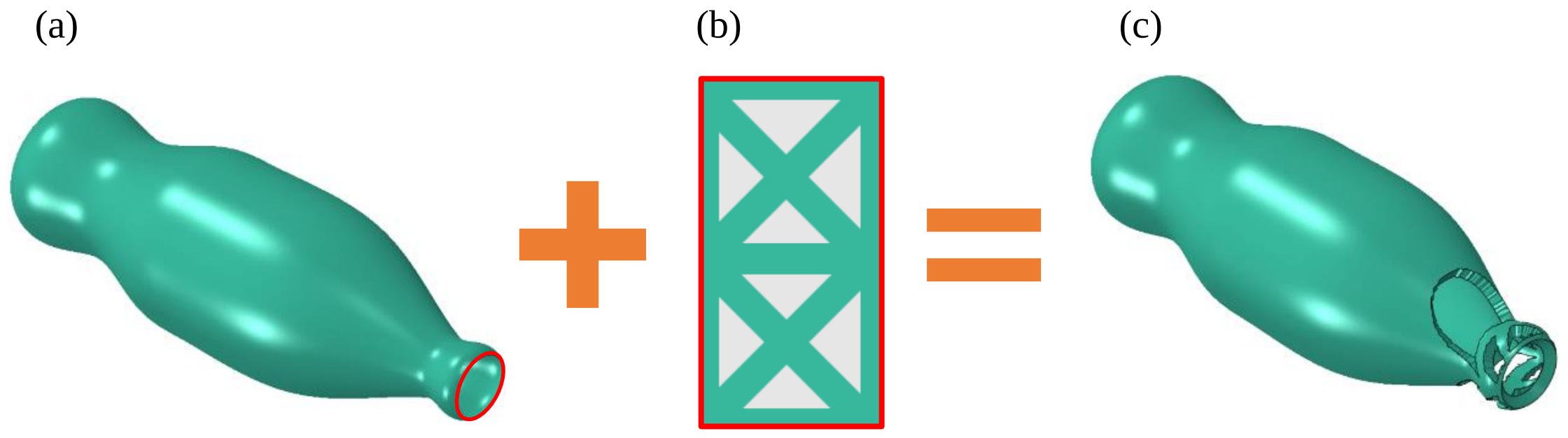

Fig. 23 Component layouts via a single-patch-based global mapping. (a) Middle surface model. (b) Components laid in the parametric domain. (c) Component layouts are severely twisted due to the global mapping process.

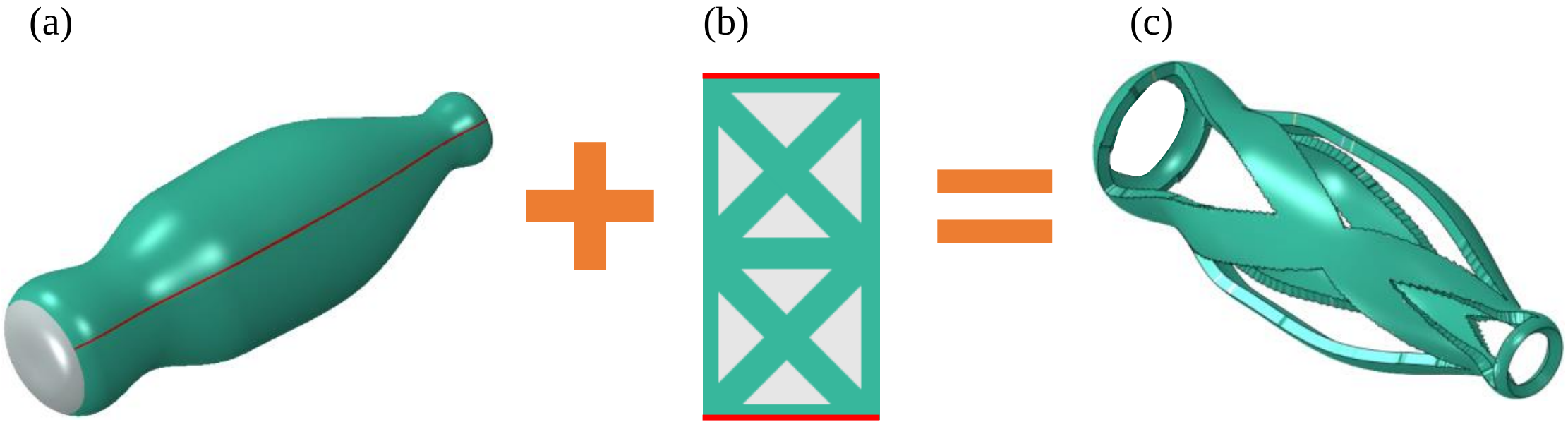

Fig. 24 Component layouts via the multi-patch stitching scheme. (a) Middle surface model. (b) Components laid in the parametric domain. (c) Component layouts preserves the fidelity of geometry modeling owing to the multi-patch stitching scheme.

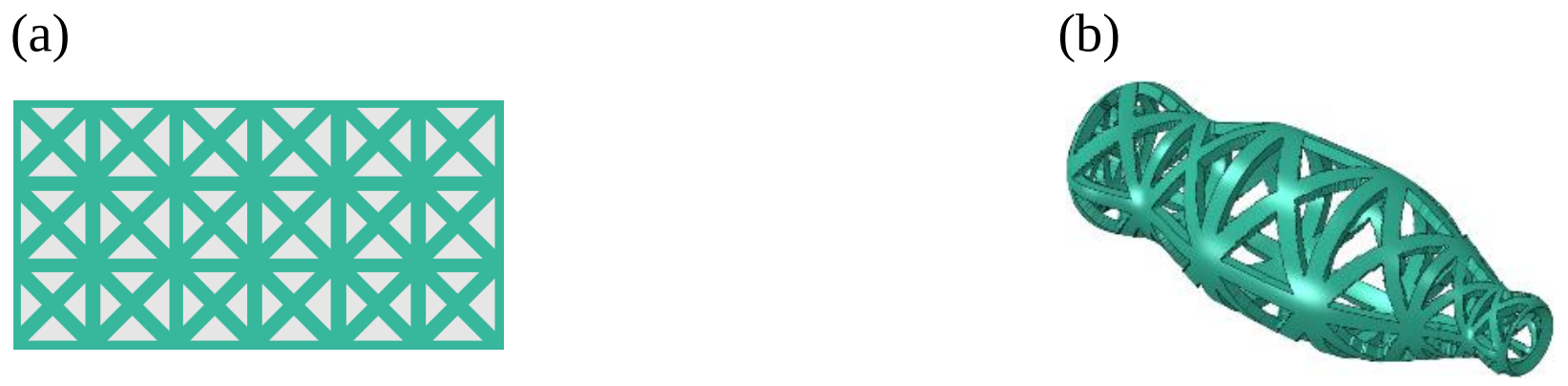

Fig. 25 Initial layout of components. (a) In the parametric domain. (b) On the original thin-walled structure.

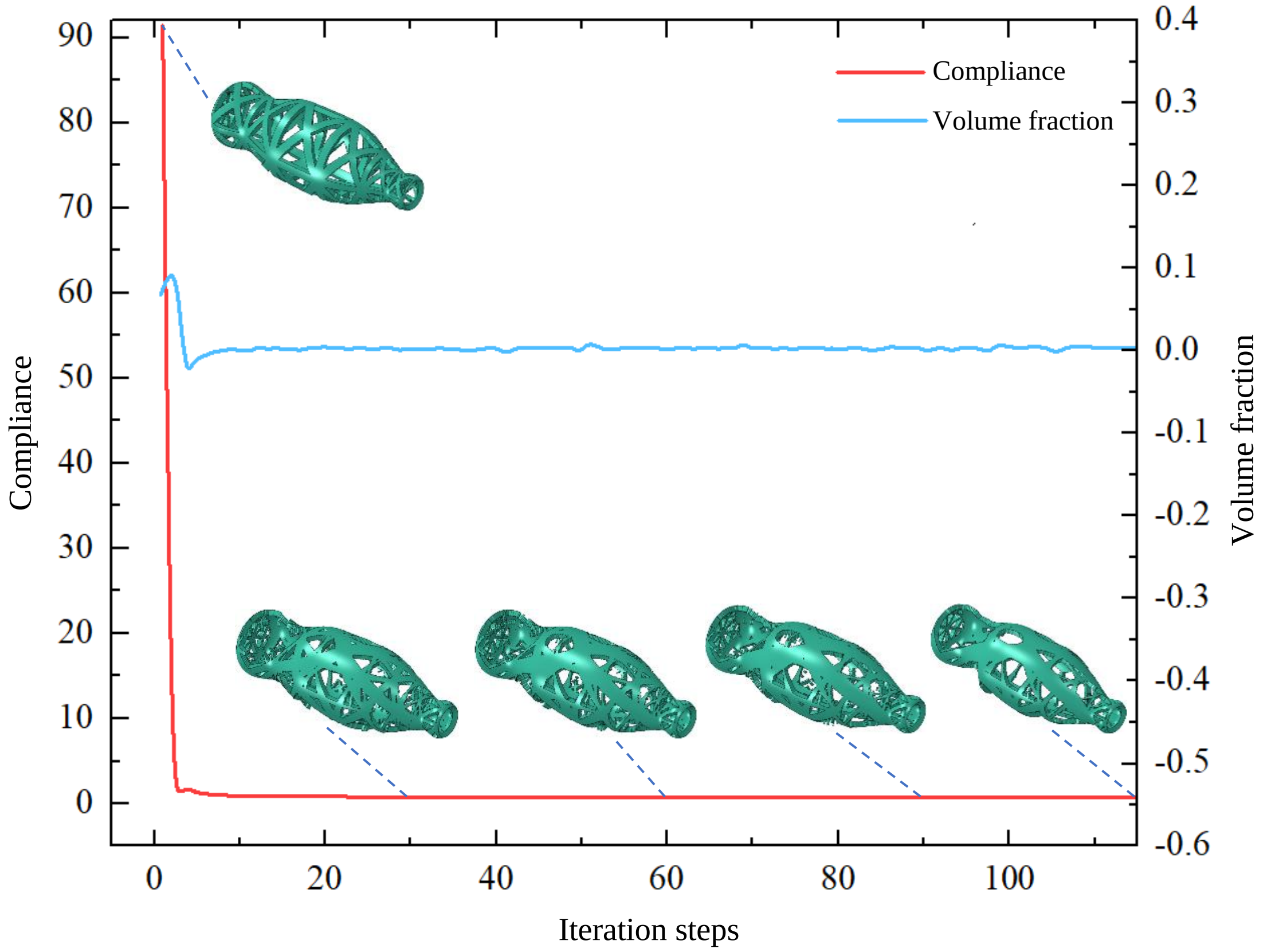

Fig. 26 Iteration histories and intermediate results of the bottle-shaped thin-walled structure.

(a) 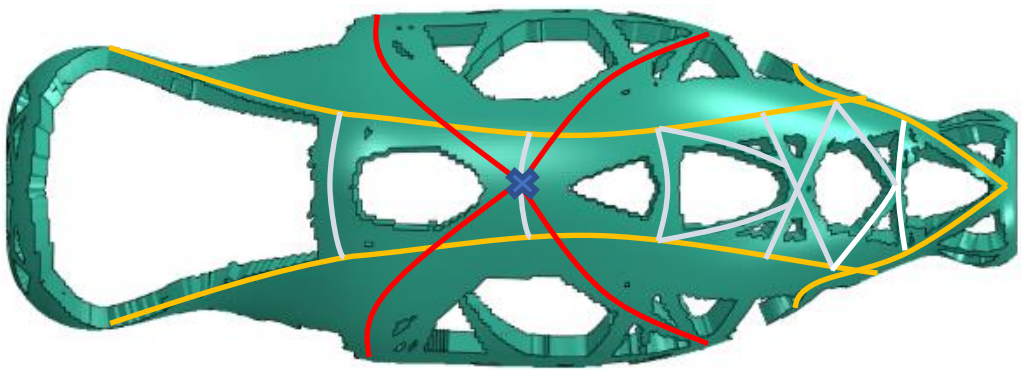

(b) 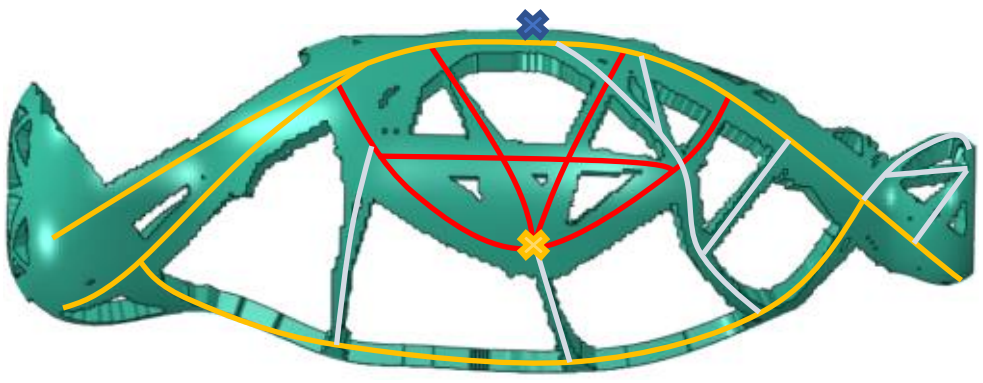

Fig. 27 Final design of the bottle example is composed of three categories of hierarchical components with different functionalities ($C^{\mathrm{opt}} = 0.6462$). (a) Top view. (b) Side view.

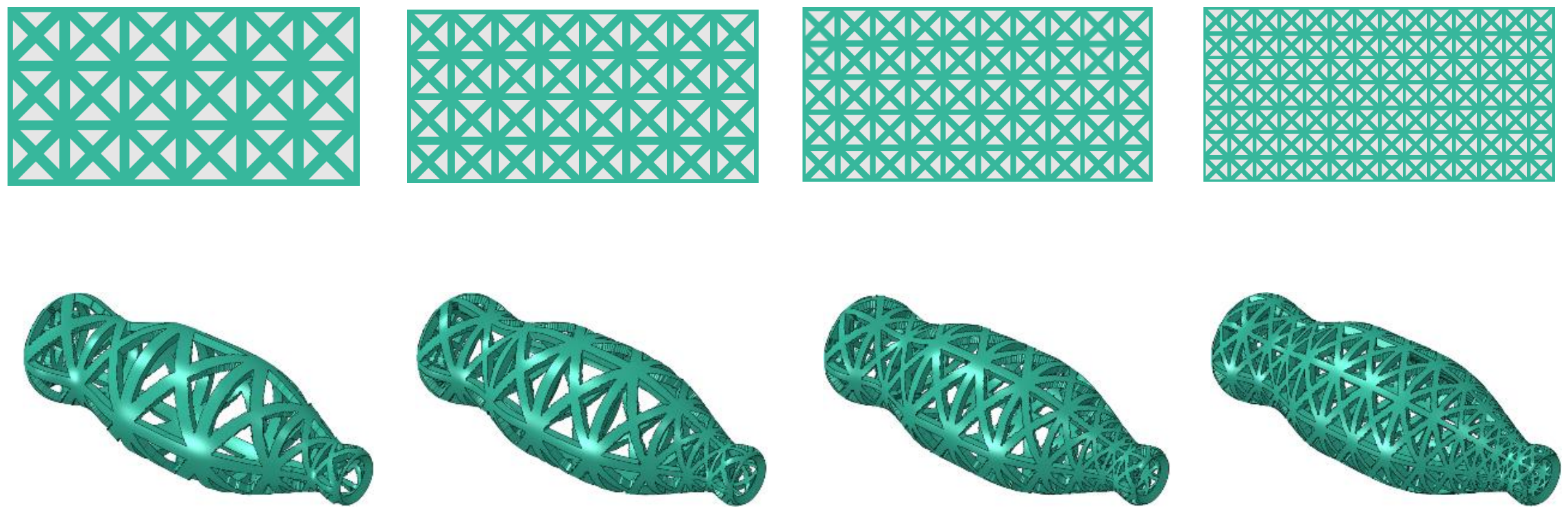

Fig. 28 Initial layouts of components with different parameter settings.

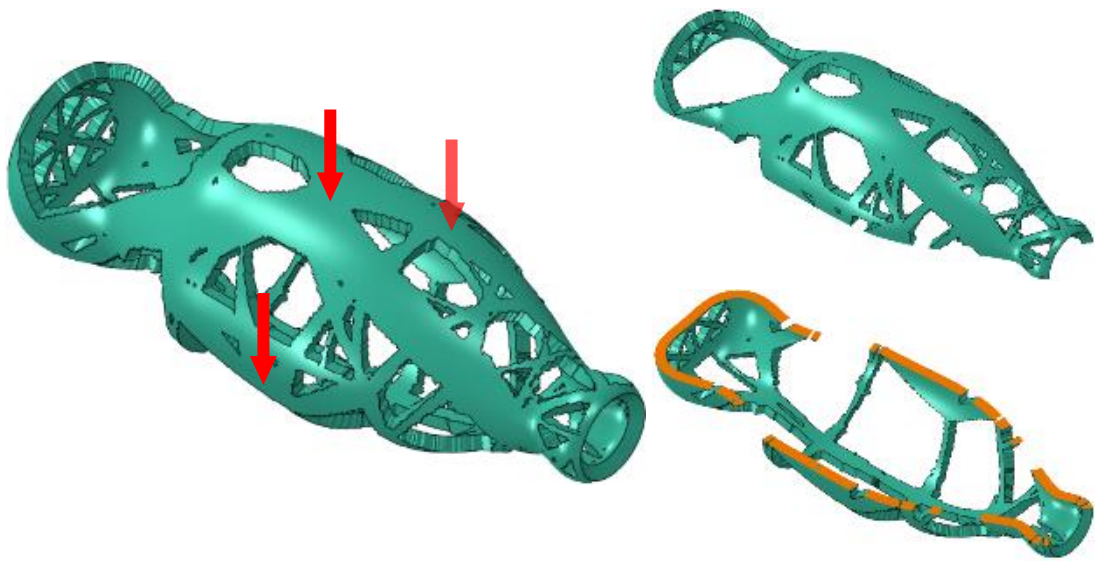

(a) $3 \times 6$ cells ($C^{\mathrm{opt}} = 0.6462$)

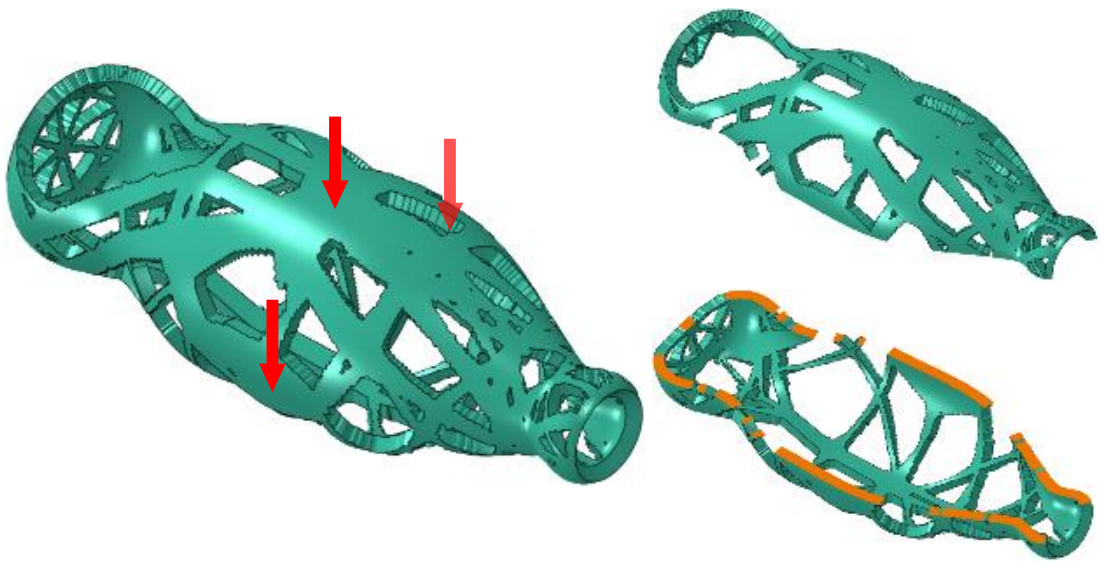

(b) $4 \times 8$ cells ($C^{\mathrm{opt}} = 0.6287$)

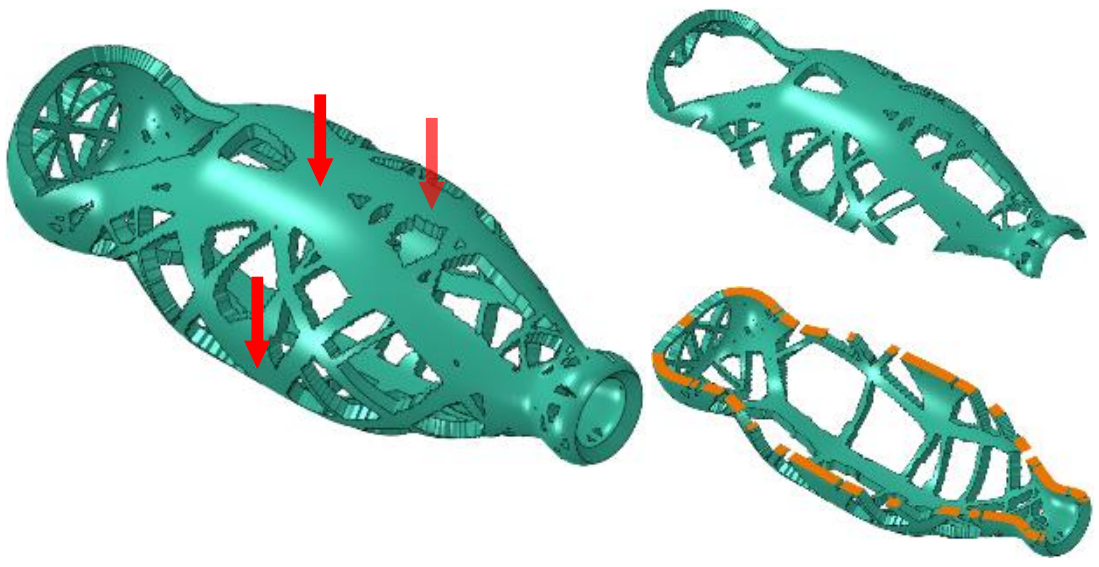

(c) $5 \times 10$ cells ($C^{\mathrm{opt}} = 0.6238$)

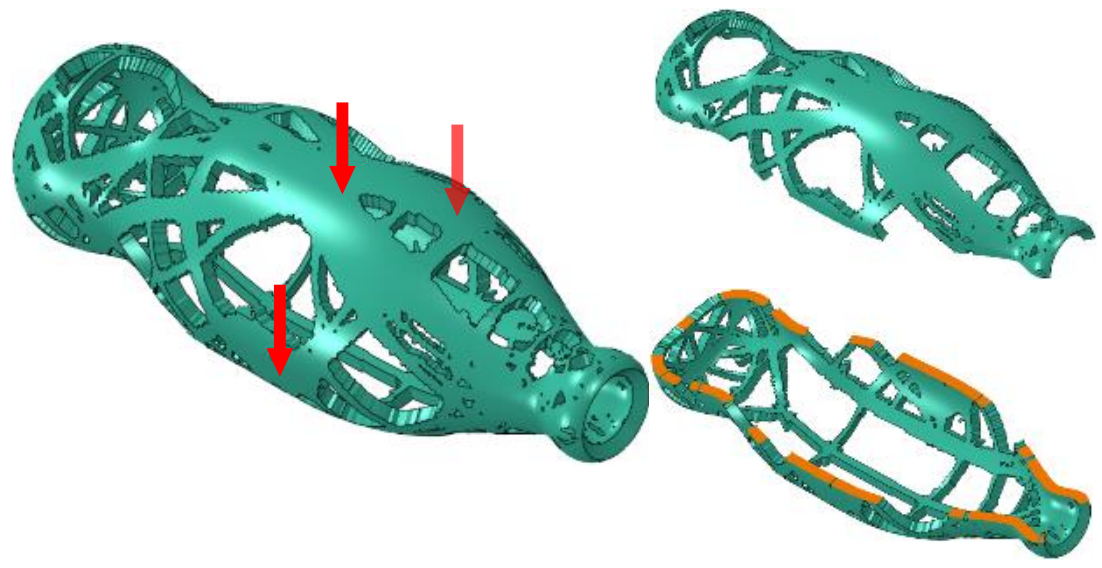

(d) $7 \times 14$ cells ($C^{\mathrm{opt}} = 0.6157$)

Fig. 29 Final designs of different initial layouts.

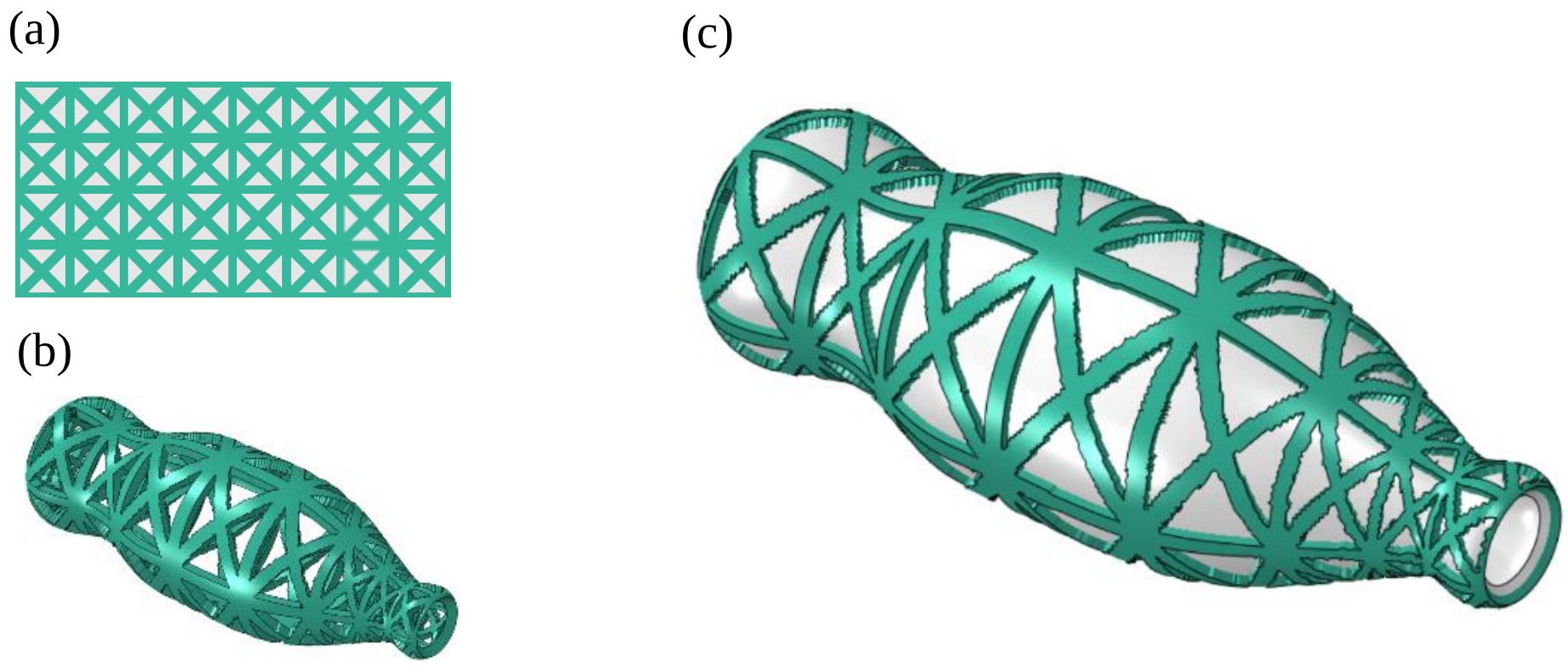

Fig. 30 Problem settings of rib-reinforced structure design. (a) and (b) The layout of components is set as $4 \times 8$ cells; (c) The thickness of the base panel is set as half of the whole thickness.

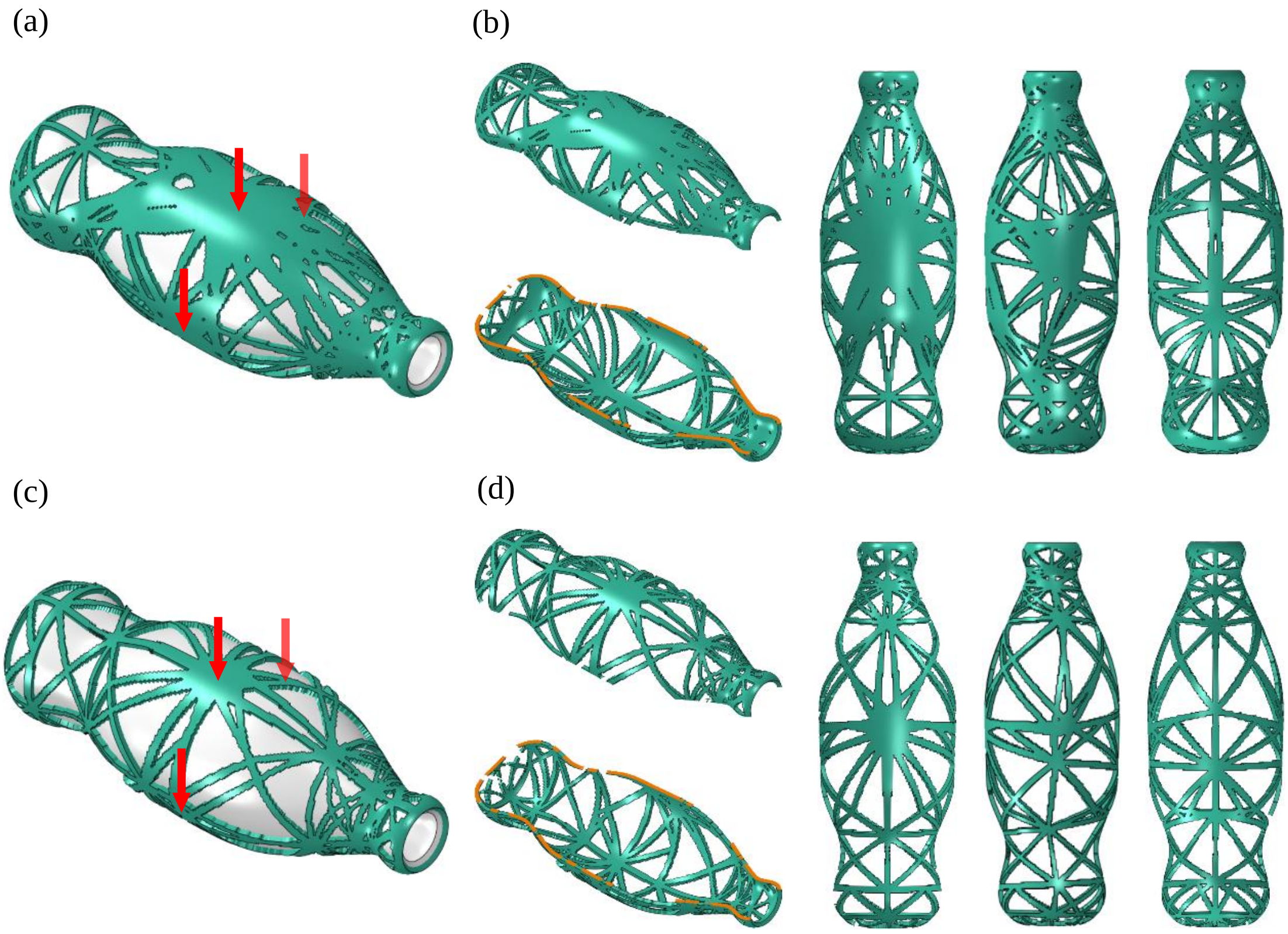

Fig. 31 Two representative designs of rib-reinforced structures. (a) Optimized result with the base panel of the first parameter setting ($C^{\mathrm{opt}} = 1.25$). (b) Optimized results without the base panel of the first parameter setting. (c) Optimized result with the base panel of the second parameter setting ($C^{\mathrm{opt}} = 0.54$). (d) Optimized results without the base panel of the second parameter setting.

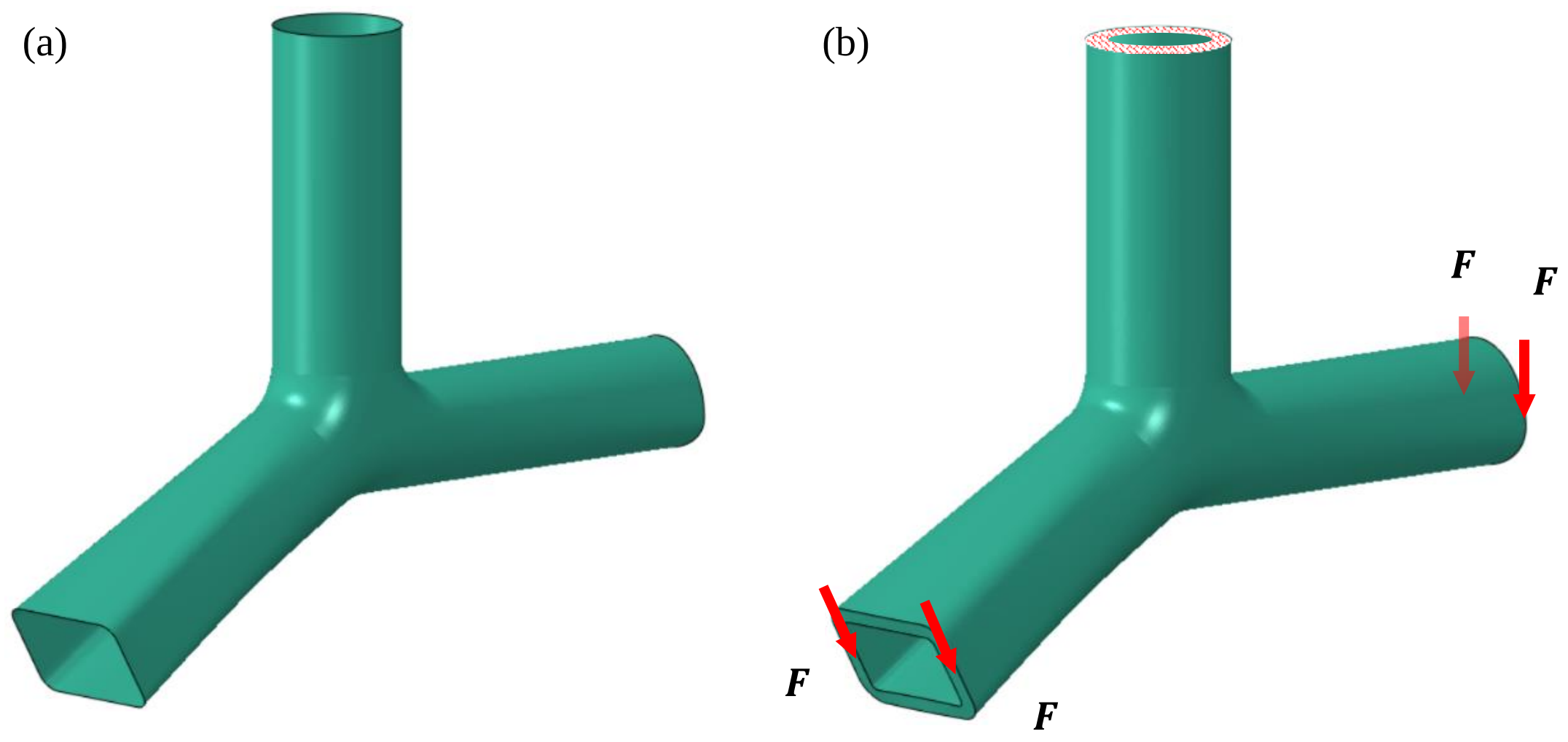

Fig. 32 Geometry and boundary conditions of the tee-branch pipe example. (a) Middle surface model. (b) Solid model and boundary conditions.

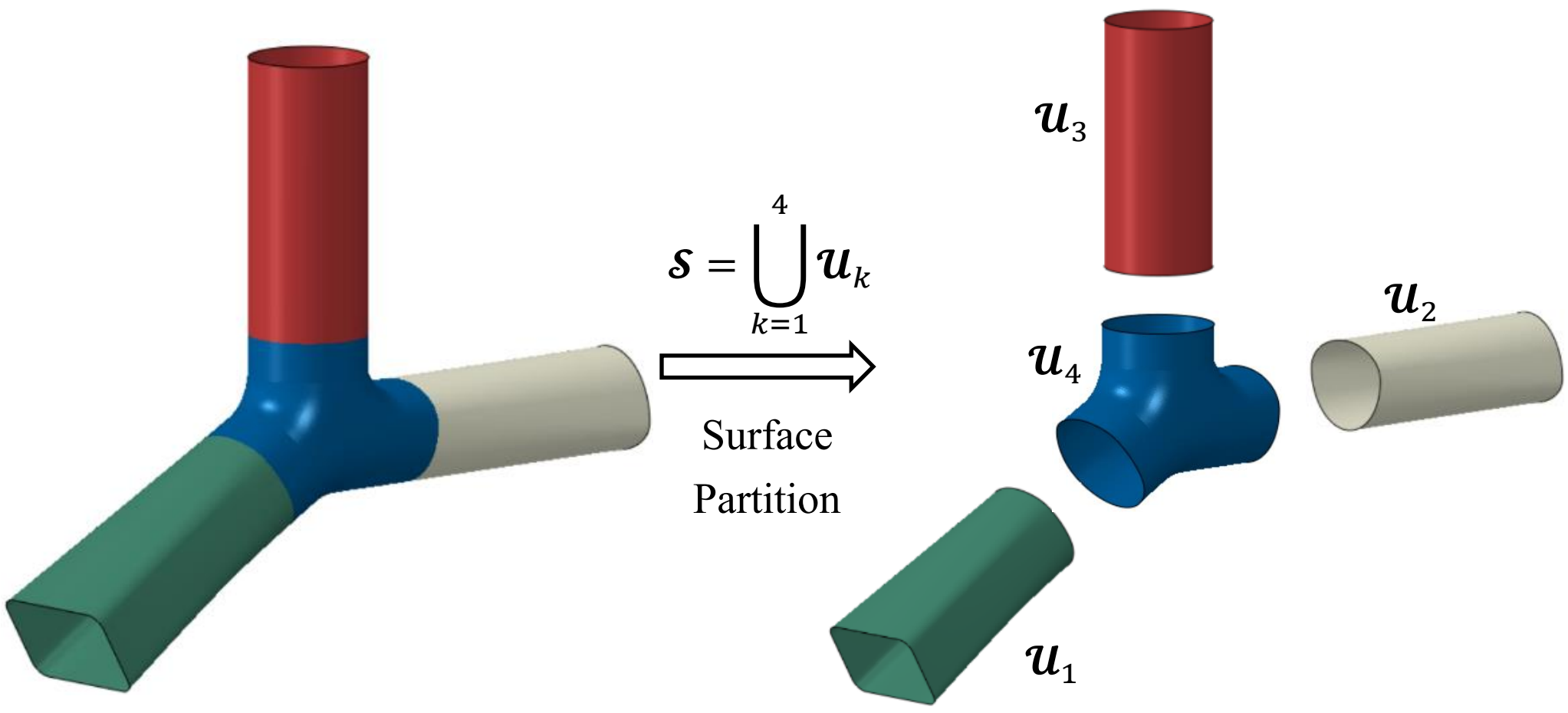

Fig. 33 Partitioning the original surface into four patches.

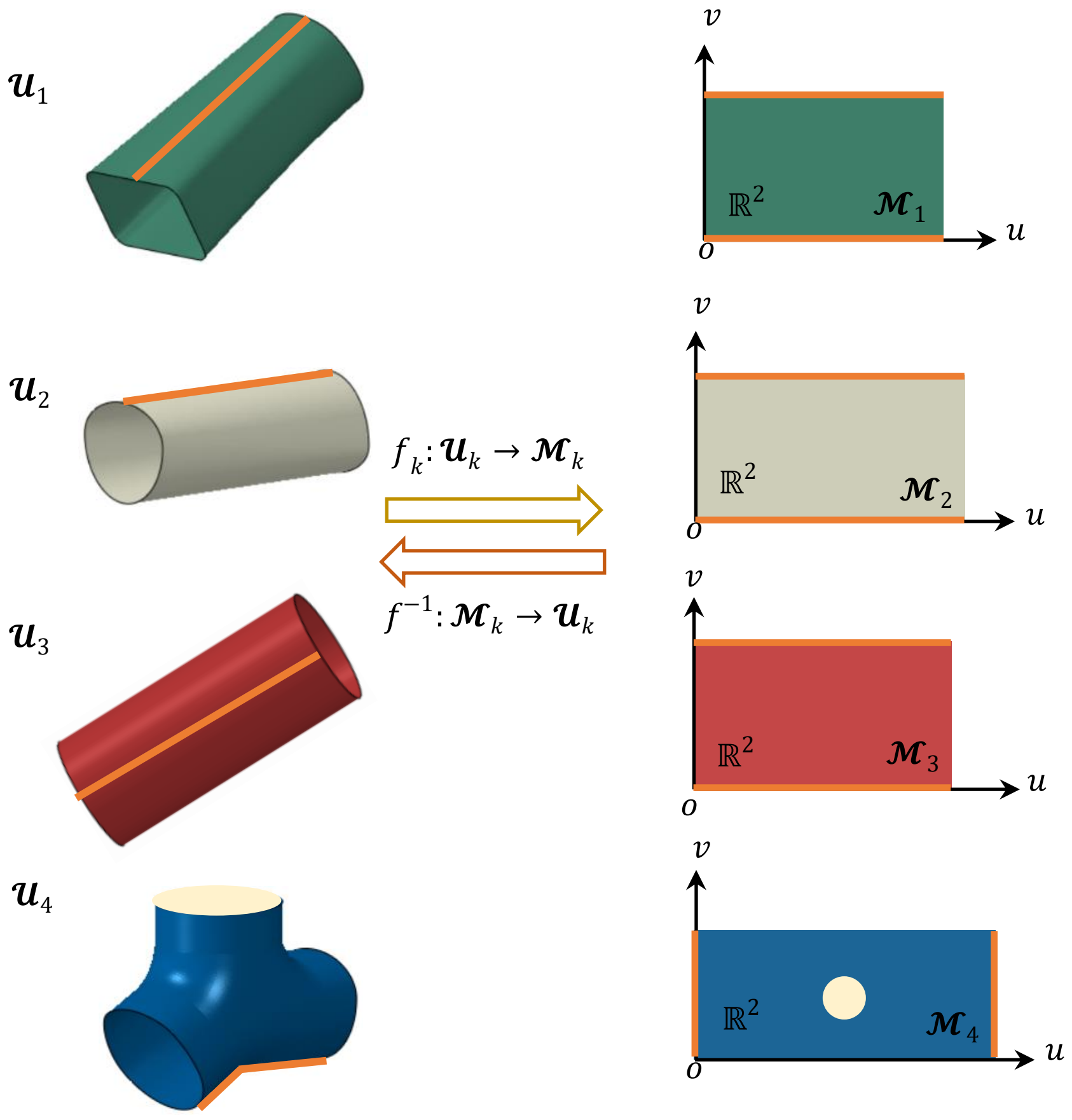

Fig. 34 Parameterization of each patch.

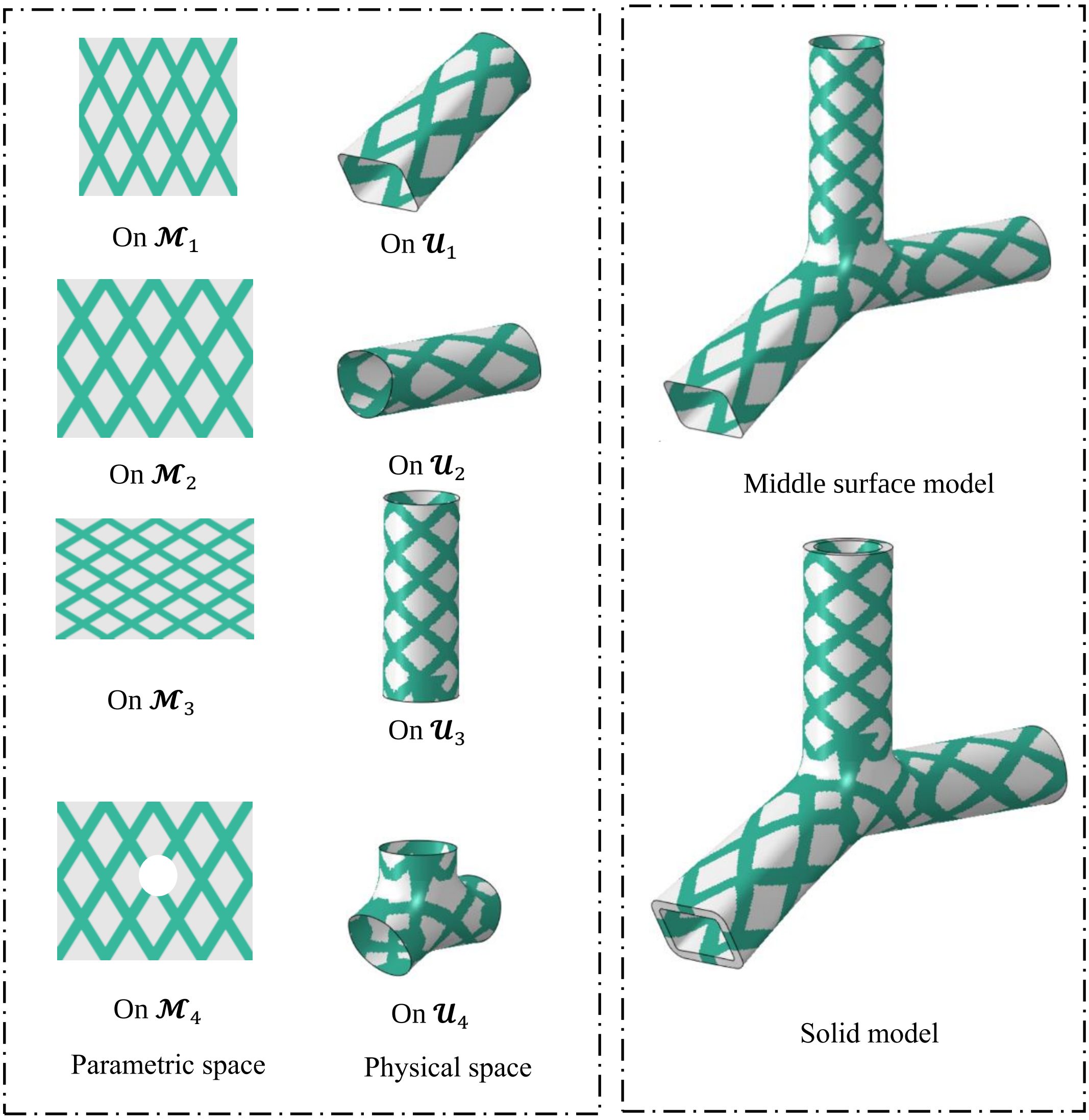

(a) Initial component layout on each patch. (b) Assembling of all patches.

Fig. 35 Initial component layout of the tee-branch pipe thin-walled structure.

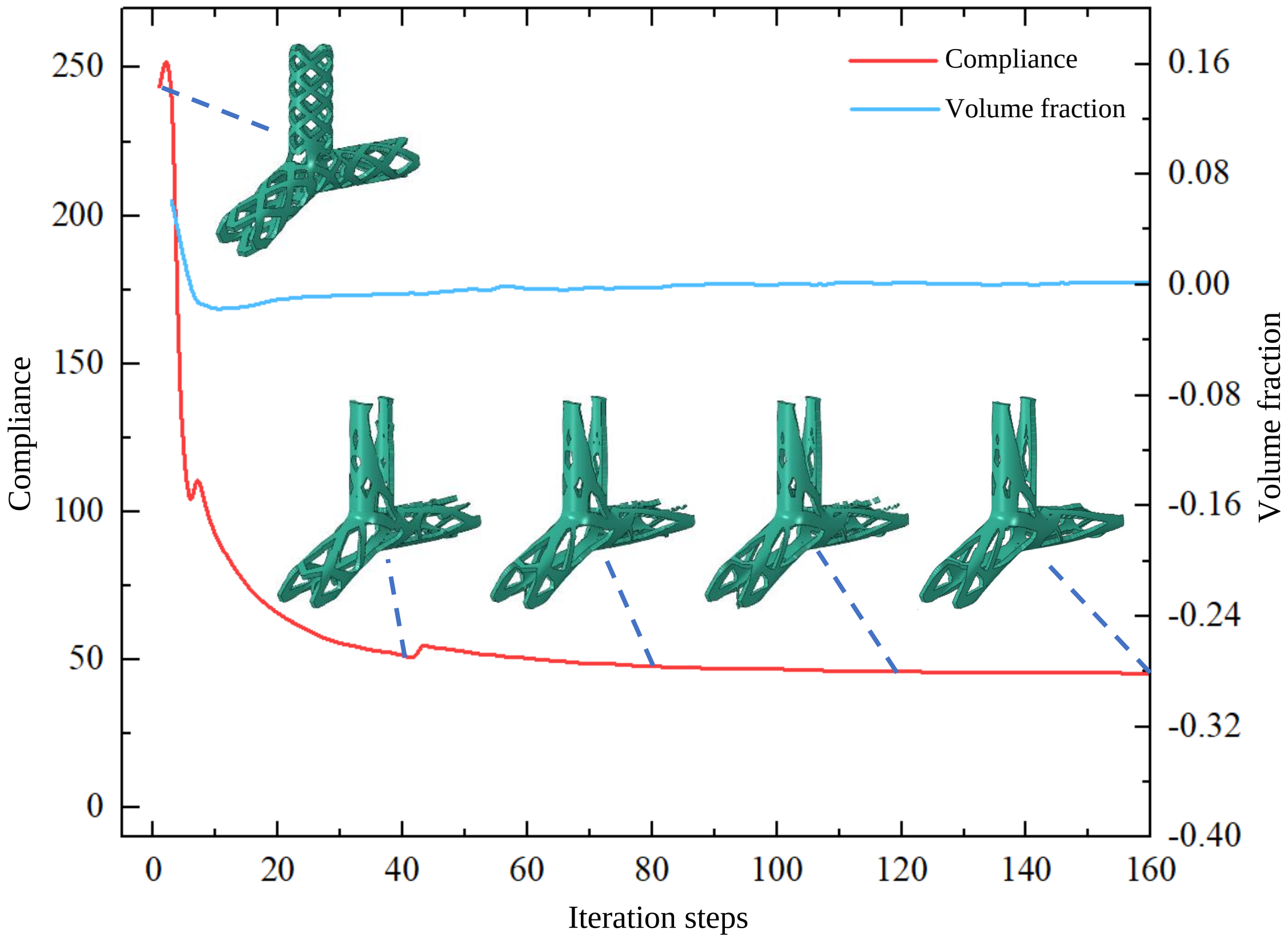

Fig. 36 Iteration histories and intermediate designs.

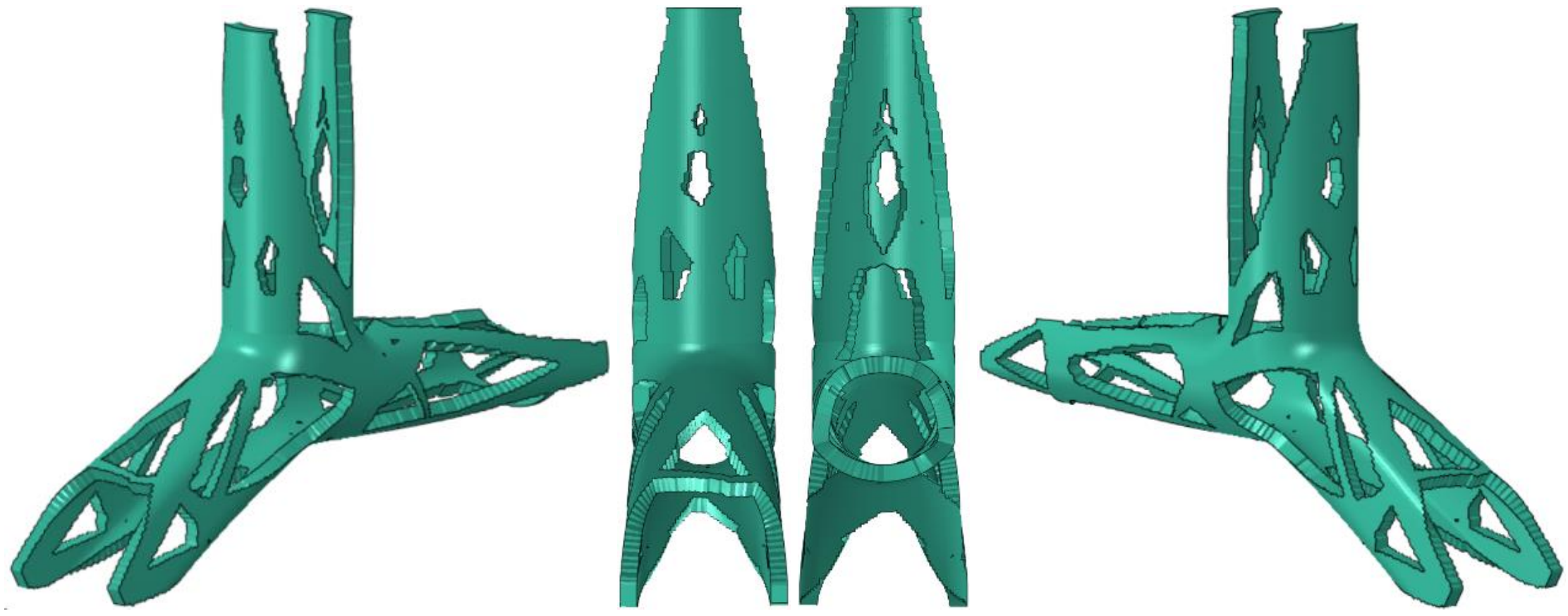

Fig. 37 Optimized result ($C^{opt} = 43.38$).

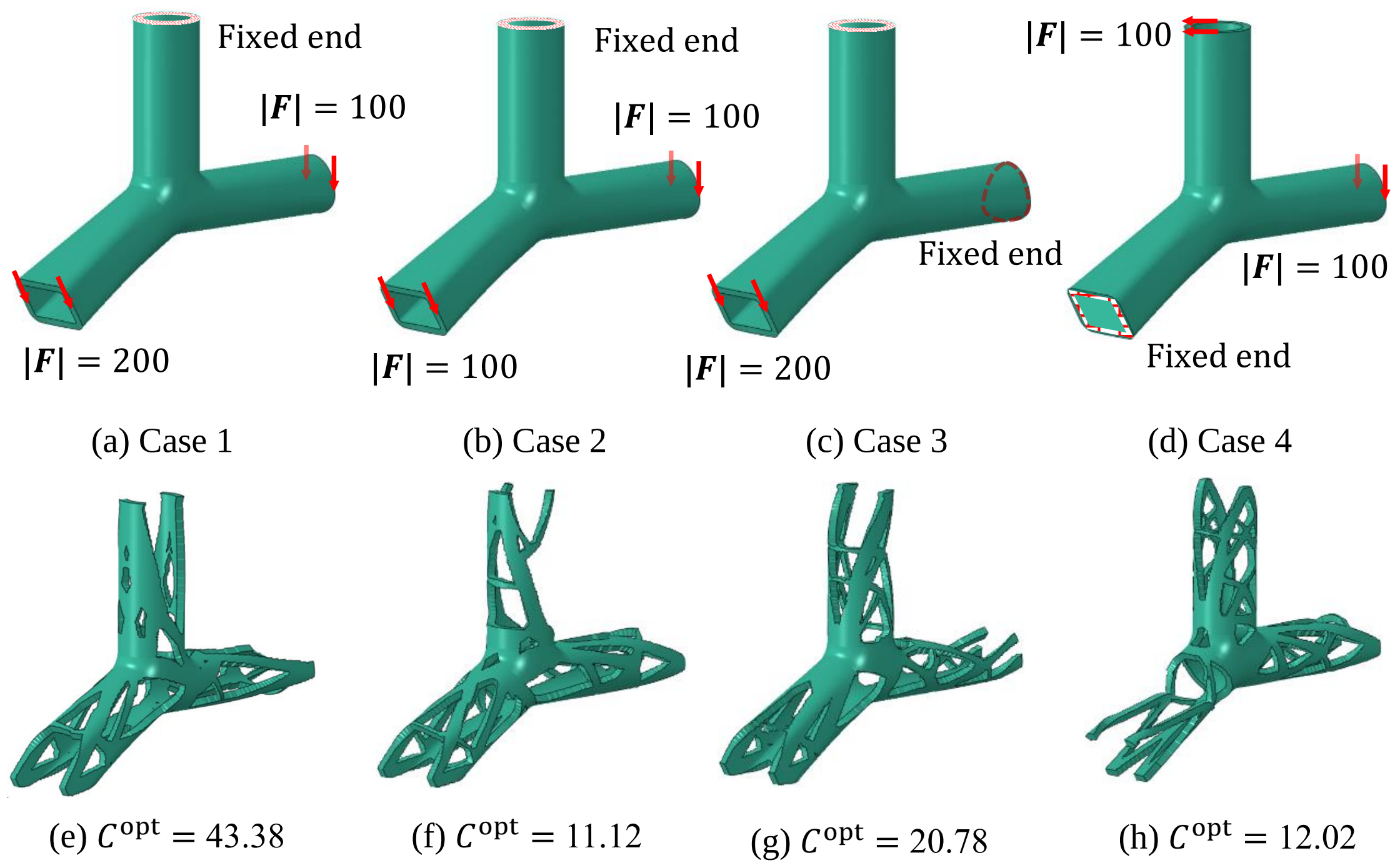

Fig. 38 Final designs with different boundary conditions.

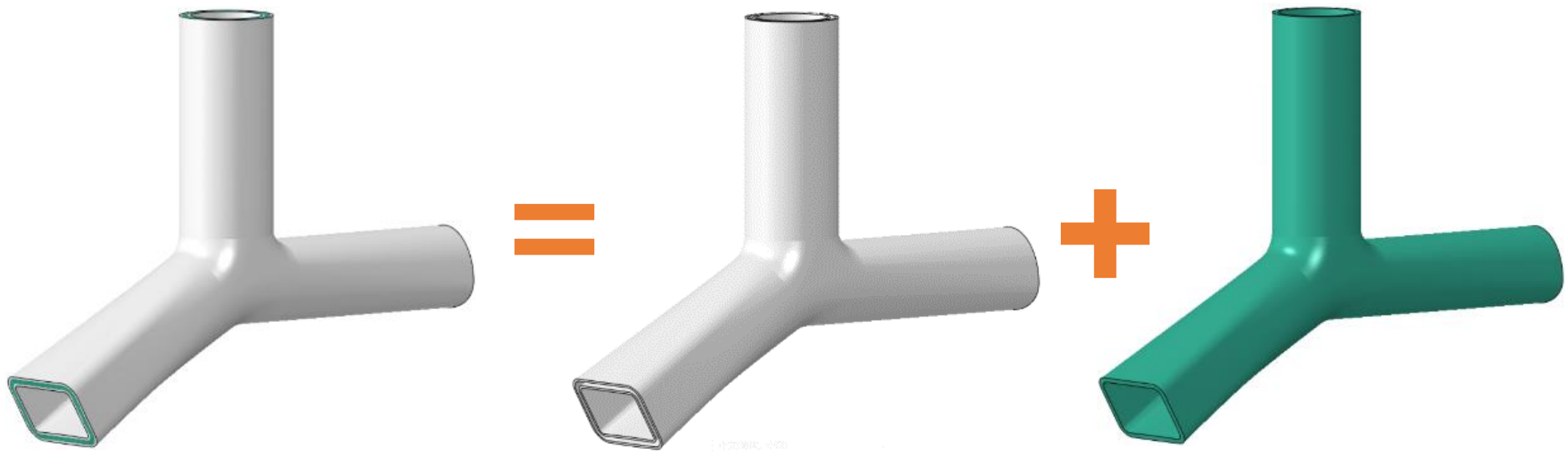

(a) Whole structure.

(b) Undesigned domain.

(c) Designed domain.

Fig. 39 Thickness settings of the sandwich-type reinforced structure design.

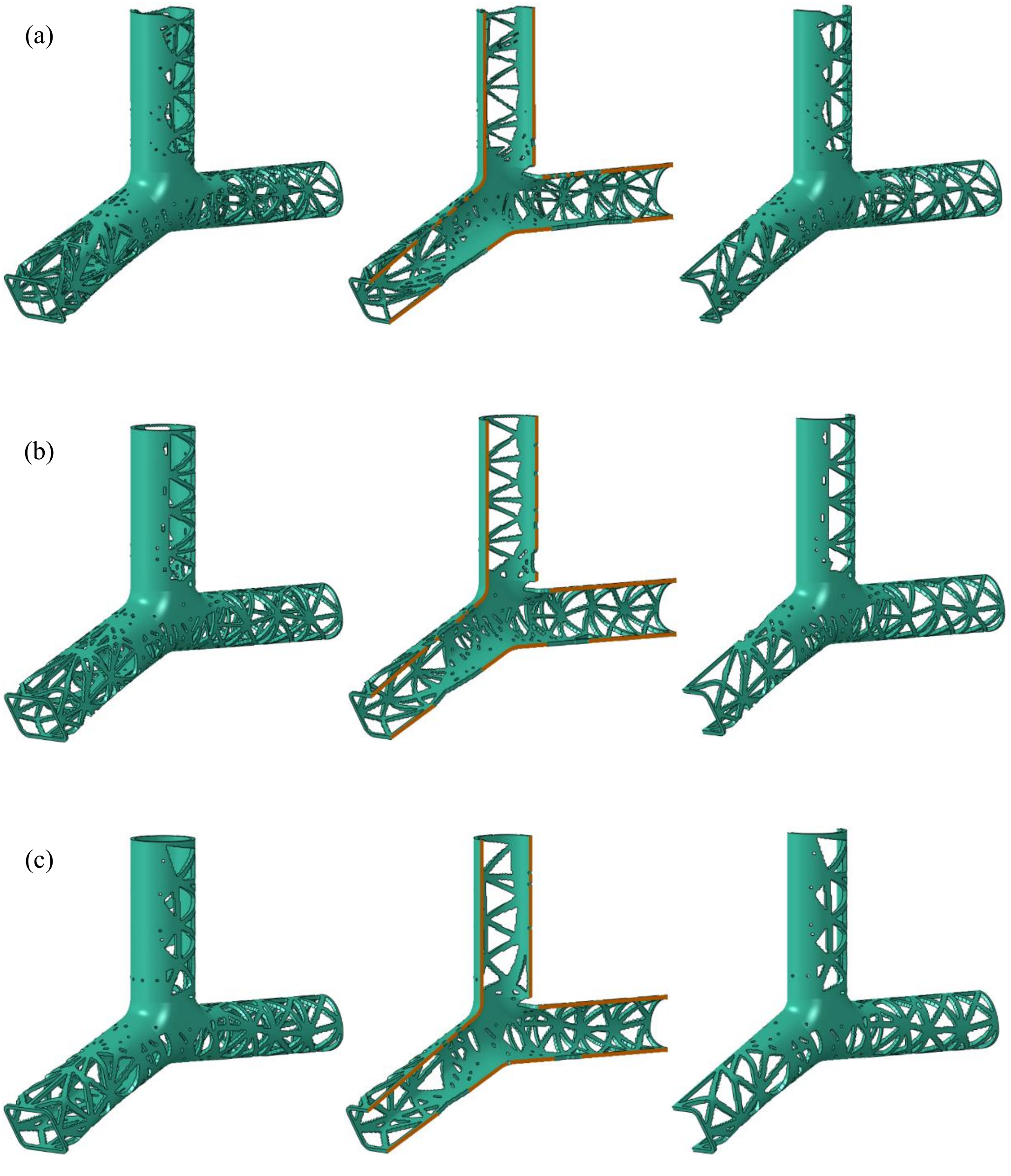

Fig. 40 Final structures of the designed domains in the sandwich-type reinforced structure. From (a) to (c), the volume fractions are set as 0.375, 0.45 and 0.55, and the Young's modulus of the undesigned layers are set as 0.15, 0.3, 0.5, respectively.